\documentclass[preprint,3p]{elsarticle}

\usepackage[toc,page,title,titletoc,header]{appendix}
\usepackage[breaklinks,colorlinks,linkcolor=black,citecolor=black,urlcolor=black]{hyperref}
\usepackage{stmaryrd}
\SetSymbolFont{stmry}{bold}{U}{stmry}{m}{n}
\usepackage{amsmath}
\usepackage{amssymb}
\usepackage{amsthm}
\usepackage{tabularx}
\usepackage{multirow}
\usepackage{epsfig}
\usepackage{cases}
\usepackage{cleveref}
\biboptions{sort&compress}
\usepackage{enumitem}
\usepackage{graphicx}
\usepackage{epsfig}
\usepackage{color}
\usepackage{mathrsfs}
\usepackage{caption}

\newcommand{\norm}[1]{\lVert#1\rVert}
\renewcommand{\vec}[1]{\mathbf{#1}}

\newcommand{\bmath}[1]{\mbox{\boldmath{$#1$}}}
\newcommand{\rd }{\mathrm{d}}
\newcommand{\nn}{\nonumber}

\newcommand{\be}{\begin{equation}}
\newcommand{\ee}{\end{equation}}
\newcommand{\ba}{\begin{array}}
\newcommand{\ea}{\end{array}}
\newcommand{\bea}{\begin{eqnarray}}
\newcommand{\eea}{\end{eqnarray}}
\newcommand{\beas}{\begin{eqnarray*}}
\newcommand{\eeas}{\end{eqnarray*}}

\newcommand{\tabincell}[2]{\begin{tabular}{@{}#1@{}}#2\end{tabular}}

\newcommand{\bt}{{\mathbf t}}
\newcommand{\bn}{{\mathbf n}}
\newcommand{\bu}{\mathbf{u}}
\newcommand{\bJ}{\mathbf{J}}

\newcommand{\ds}{\mathrm{d}s}

\newcommand{\dt}{\mathrm{d}t}

\newcommand{\rD}{\rm D}

\newcommand{\dH}{\mathrm{d}\mathscr{H}}

\newtheorem{thm}{Theorem}[section]

\numberwithin{equation}{section}

\crefname{section}{Section}{Section}

\begin{document}

\begin{frontmatter}

\title{A Thermodynamically Consistent Model and Its Conservative Numerical Approximation for Moving Contact Lines with Soluble Surfactants}

\author[1]{Quan Zhao}
\ead{matzq@nus.edu.sg}
\address[1]{Department of Mathematics, National University of Singapore, Singapore, 119076}

\author[1]{Weiqing Ren}
\ead{matrw@nus.edu.sg}

\author[2]{Zhen Zhang\corref{corad}}
\ead{zhangz@sustech.edu.cn}
\address[2]{Department of Mathematics,
Guangdong Provincial Key Laboratory of Computational Science and Material Design, Southern University of Science and Technology, Shenzhen, Guangdong, China, 518055}

\cortext[corad]{Corresponding author.}


\begin{abstract}
We derive a continuum sharp-interface model for moving contact lines with soluble surfactants in a thermodynamically consistent framework. The model consists of the isothermal two-phase incompressible Navier-Stokes equations for the fluid dynamic and the bulk\slash surface convection-diffusion equations for the surfactant transportation. The interface condition, the slip boundary condition, the dynamic contact angle condition, and the adsorption\slash desorption condition are derived based on the principle of the total free energy dissipation. In particular, we recover classical adsorption isotherms from different forms of the surface free energy. The model is then numerically solved in two spatial dimensions. We present an Eulerian weak formulation for the Navier-Stokes equations together with an arbitrary Lagrangian-Eulerian weak formulation for the surfactant transport equations. Finite element approximations are proposed to discretize the two weak formulations on the moving mesh. The resulting numerical method is shown to conserve the total mass of the surfactants exactly. By using the proposed model and its numerical method, we investigate the droplet spreading and migration in the presence of surfactants and study their dependencies on various dimensionless adsorption parameters.
\end{abstract}



\end{frontmatter}
\section{Introduction}

Surfactants are surface-active substances that decrease the surface tension of the interface between two fluids or two phases of one fluid. They contain hydrophobic tails and hydrophilic heads, and thus can be adsorbed from the bulk fluid to the interface. Surfactants have found wide industrial applications as cleaning detergents, emulsifiers, dispersants, foaming and anti-foaming agents, and scientific applications in micro-fluidics~\cite{Eggleton2001tip,Branger2002,Baret2012}. The presence of surfactants in the multi-phase system has great effects on the dynamics of the interface by modifying the interfacial capillary force. Besides, a nonuniform distribution of surfactants along the interface can lead to a gradient of the surface tension and thus induce a tangential Marangoni force. The dynamics of the three-phase intersection, such as contact line motion, is also influenced by the surfactant transport. Modeling and simulation for moving contact lines with surfactants have attracted much attention in recent years.

\begin{figure}[t]
\center
\includegraphics[width = 0.6\textwidth]{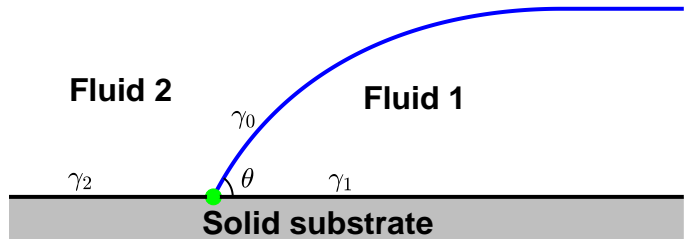}
\caption{Two immiscible fluids in contact with a solid-substrate, where $\gamma_0$ is the surface tension coefficient of the fluid interface, and $\gamma_1$, $\gamma_2$ are surface tension coefficients of the fluid-solid interfaces.}
\label{fig:equilibrium}
\end{figure}

When two immiscible fluids are placed on a solid substrate, a moving contact line (MCL) forms at the intersection of the fluid interface and the solid substrate. The static contact angle $\theta_Y$ of the interface satisfies the Young-Dupr\'e equation (see Fig.~\ref{fig:equilibrium})
\begin{equation}
\gamma_0\cos\theta_Y = \gamma_2-\gamma_1,
\end{equation}
which prescribes a balance of the tangential components among the fluid-fluid surface tension $\gamma_0$ and the two fluid-solid surface tensions $\gamma_1$ and $\gamma_2$. The MCL problem has attracted a lot of attention in recent decades. The main difficulty in this problem arises from the stress singularity at the MCL in classical hydrodynamic models with the conventional no-slip boundary condition \cite{Huh71, Dussan74}. A lot of efforts have been devoted to resolving this difficulty, and
different models have been proposed. These include molecular dynamics models
\cite{Koplik88, Thompson89, Ren07, DeConinck08},
diffuse interface models \citep{Anderson98, Jacqmin00, Pismen02, Qian03, Yue10},
and hydrodynamic models
\cite{Voinov76, Hocking77, Cox86, Eggers04a, Ren10, Ren11d, Ren15, ZhangRen2019, Sibley15}. We refer to the review articles~\cite{Dussan79, deGennes85, Kistler93, Pomeau02, Bonn09},
the collected volume \cite{Velarde11}
and the monographs \cite{deGennes03, Starov07}
for detailed discussions of the MCL problem.

In those models for MCL, the thermodynamic principle plays an important role. In the continuum sharp-interface hydrodynamic model developed in~\cite{Ren07,Ren10,Ren11}, the fluid dynamic is modeled by the isothermal incompressible Navier-Stokes equations and the rate of change in the total free energy is considered. The slip boundary condition and the contact angle condition are derived in order to guarantee that the total free energy decays with time, which is a result of the second law of thermodynamics. The detailed constitutive relations in these boundary conditions can be computed from molecular dynamics simulations. The underlying thermodynamic principle has another form which is usually called the Onsager variational principle \cite{Doi13}. The Onsager principle has become a useful tool in the derivation of many MCL models \cite{qian06variational, xu16variational}.

In the presence of surfactants, the fluid-fluid surface tension is modified according to the surfactant concentration on the interface. As a result, the contact angle and the contact line dynamics are influenced. In an ideal case of insoluble surfactants, the surfactants are not allowed to dissolve in the bulk fluid but can only stay on the fluid-fluid interface. The mass conserved convection-diffusion equation on the surface was derived for the transportation of the insoluble surfactants \cite{stone90simple, wong96surfactant}. Taking into account the surfactant-dependency of the surface free energy, Zhang et al. \cite{Zhang14} derived a thermodynamically consistent sharp-interface model for the contact line dynamics with insoluble surfactants and obtained the energy dissipation law.
In the phase-field framework, Zhu et al. \cite{Zhu2019} considered the contribution of soluble surfactants in the total free energy and developed a diffuse-interface model for moving contact lines based on the Onsager principle. We also refer to \cite{Garcke2014diffuse, thiele16gradient} for more discussions of the thermodynamically consistent modeling of multiphase flows with surfactants.

There exist a lot of numerical work in the literature for the simulation of two-phase flows with surfactants. These include volume of fluid (VOF) method \cite{Renardy2002new, James04surfactant, Alke20093d, Afkhami09}, level set method \cite{Xu2018level, Titta2018level}, phase field approach \cite{Liu2010phase, Teigen2011diffuse, Garcke2014diffuse}, immersed boundary method \cite{lai08immersed, Chen2014}, and other related work \cite{Zhang2006front, Muradoglu2008front, Ganesan12, Booty2010hybrid, Bazhlekov2006,Khatri2011,Barrett2015insoluble, Barrett2015soluble}.
The main difficulty in the numerical methods arises in the capture of the free boundary and the computation of the surfactant concentration with satisfactory accuracy. The numerical preservation of the surfactant mass conservation and the energy dissipation law is also of particular interest. This situation becomes more complicated when there is a moving contact line. A good numerical treatment of the slip boundary condition and the contact angle condition becomes important. Some efforts have been made in the numerical simulation of moving contact lines with surfactants \cite{Lai2010numerical, Xu14, Zhu2019}.

This work is devoted to the modeling and numerical study of moving contact lines with soluble surfactants. We extend the model derivation in \cite{Ren11, Ren11d, Zhang14} and derive a continuum sharp-interface model for the moving contact lines with soluble surfactants. The dynamic of the soluble surfactants is modeled by a coupled system of bulk\slash surface convection-diffusion equations. The total mass conservation of the surfactants is maintained by the no-flux boundary condition on the solid substrate and the adsorption\slash desorption condition on the interface between the bulk and interfacial surfactants. In account of surfactants, we add the contribution from the bulk and interfacial surfactants into the total free energy. In the presence of two-phase incompressible Navier-Stokes flow, we obtain the energy dissipation law and derive the surfactant-corrected contact angle condition together with the adsorption\slash desorption rate. In particular, by taking different forms of the surfactant-dependent surface free energy, we recover the classical adsorption isotherms.

Based on the recent numerical work \cite{Zhao19, Zhao20ewod} for moving contact lines, we develop a numerical method for the model in the finite element framework. Due to the variational structure of the model, two weak formulations are immediately available: (i) an Eulerian weak formulation for the two-phase incompressible Naiver-Stokes system; and (ii) an arbitrary Lagrangian-Eulerian (ALE) weak formulation for the surfactant transport equations. The moving mesh approach is used so that the bulk mesh remains fitted to the evolving fluid interface. We numerically discretize the two weak formulations on the moving mesh and obtain decoupled linear systems. The numerical method maintains a satisfactory mesh quality by introducing an implicit tangential velocity for the markers on the fluid interface. In addition, the numerical method conserves the total mass of surfactants exactly.

The rest of the paper is organized as follows. We begin in \cref{sec:DerivationModle} with the derivation of the continuum model based on thermodynamic principles and conservation laws. In  \cref{sec:dmodel}, we consider the dynamic system in a bounded domain in two spatial dimensions and propose the weak form of the dimensionless model. In \cref{se:FEM}, we present the finite element approximations based on the weak form. We then report the convergence test and other numerical experiments in \cref{sec:numer} and finally draw a conclusion in \cref{sec:con}.

\section{The mathematical model}
\label{sec:DerivationModle}
\begin{figure}[!htp]
\centering
\includegraphics[width=0.8\textwidth]{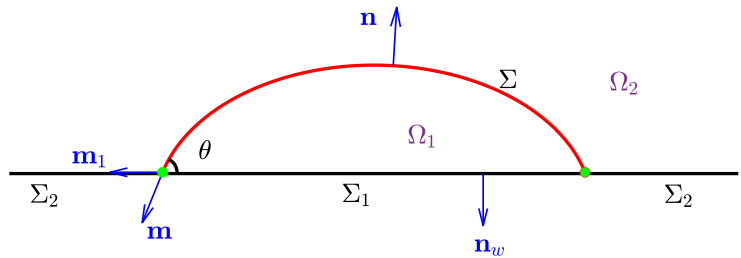}
\caption{An illustration of a droplet sitting on a flat solid substrate. }
\label{Fig:dynamicmodel}
\end{figure}

We consider the dynamics of a droplet sitting on a flat substrate in $\mathbb{R}^d$ ($d=2,3$). As shown in Fig.~\ref{Fig:dynamicmodel}, the regions occupied by the droplet and the fluid outside the droplet are denoted by $\Omega_1$ and $\Omega_2$, respectively. The fluid interface is denoted by $\Sigma$ with the unit normal vector $\bn$ pointing to $\Omega_2$, and the fluid-solid interfaces are denote dy $\Sigma_1\cup\Sigma_2$ with the unit normal vector $\bn_w$. The fluid interface intersects with the solid substrate at the contact line denoted by $\Lambda$. $\vec m$ and $\vec m_1$ are the outward unit conormal vectors of $\Sigma$ and $\Sigma_1$ at the contact line, respectively.

In the present study, we assume that the surfactants are only soluble in the droplet i.e., $\Omega_1(t)$. The fluid interface $\Sigma(t)$ is covered by surfactants with concentration denoted by $\Gamma(\vec x,~t):\Sigma(t)\times[0,~T]\to\mathbb{R}$; while the concentration in the bulk is given by $\Phi(\vec x,~t):\Omega_1(t)\times[0,~T]\to\mathbb{R}$. Then the total free energy of the system is given by
\begin{equation}
\label{eq:Tenergy}
  W=\sum_{i=1,2}\int_{\Omega_i(t)}\frac12\rho_i|\bu|^2\dH^d+\int_{\Omega_1(t)}f(\Phi)\dH^d+\int_{\Sigma(t)}g(\Gamma)\dH^{d-1}+\int_{\Sigma_1(t)}(\gamma_1-\gamma_2)\dH^{d-1}.
\end{equation}
Here the first term is the fluid kinetic energy with $\rho_i\,(i=1,2)$ and $\bu$ being the density and velocity of the fluids, respectively; the second term represents the chemical energy due to the soluble surfactants in the bulk; the last two terms represent the surface energies of the fluid-fluid and the fluid-solid interfaces, respectively. $f$ and $g$ are two smooth functions to be given later, and $\dH^l$ represents the $l$-dimensional Hausdorff measure. Note that $\dH^3$, $\dH^2$ and $\dH^l$ and $\dH^0$ corresponds to the measures of volume, area, length, and counting respectively.

We assume the fluids are isothermal and incompressible. The fluid dynamic is then governed by the
standard Navier-Stokes equation and the incompressibility condition in the bulk $\Omega_i(i=1,2)$ as
\begin{subequations}
\label{eqn:fluiddynamic}
\begin{align}
\label{eq:fluiddynamic1}
  \rho_i\left(\frac{\partial\bu}{\partial t}+(\bu\cdot\nabla)\bu\right)&=\nabla\cdot(-p\mathbf{I}+\sigma_i),\\
  \nabla\cdot\bu&=0,
  \label{eq:fluiddynamic2}
\end{align}
\end{subequations}
where $p$ is the pressure, $\mathbf{I}\in\mathbb{R}^{{d}\times{d}}$ is the identity tensor (matrix), $\sigma_i=2\eta_i D(\vec u)$ is the viscous stress tensor with  $D(\vec u)=\frac{1}{2}(\nabla\vec u + (\nabla\vec u)^T)$ being the strain rate and $\eta_i\ (i=1,2)$ being the viscosities of the fluids.

\subsection{Mass conservation of the surfactants}
The bulk surfactant concentration $\Phi(\vec x, t)$ follows the convection-diffusion equation, which can be derived from the conservation law in any domain parcel $\hat{\Omega}\subset\Omega_1$:
\begin{equation}
\label{eqn:dcon}
  \frac{\mathrm{d}}{\dt}\int_{\hat{\Omega}}\Phi\,\dH^d+\int_{\partial\hat{\Omega}}\bJ_{_\Phi}\cdot\hat{\bn}\,\dH^{d-1}=0,
\end{equation}
where $\bJ_{_\Phi}$ is the diffusion flux and $\hat{\bn}$ the outward unit normal vector of $\partial\hat{\Omega}$. Applying the transport formula in~\eqref{eqn:dderivative} and the divergence theorem,  the left hand side of \eqref{eqn:dcon} can be recast as
\begin{equation*}
  \int_{\hat{\Omega}}\left(\frac{\rD \Phi}{\rD t} + \Phi\,\nabla\cdot\bu\right)\,\dH^d+\int_{\partial\hat{\Omega}}(\bJ_{_\Phi}\cdot\bn)\,\dH^{d-1}=\int_{\hat{\Omega}}\left(\frac{\partial \Phi}{\partial t}+\nabla\cdot(\Phi\bu)+\nabla\cdot\bJ_{_\Phi}\right)\,\dH^d,
\end{equation*}
where we define the material derivative $\frac{\rD\Phi}{\rD t}$ via \eqref{eqn:mderivative}.
Since the above equation holds in any $\hat{\Omega}\subset\Omega_1$, we arrive at the differential form of the conservation law for the surfactants in the bulk domain $\Omega_1(t)$,
\begin{equation}
\label{eqn:bsdynamic}
  \frac{\partial \Phi}{\partial t}+\nabla\cdot(\Phi\bu) + \nabla\cdot\bJ_{_\Phi}=0.
\end{equation}
In the case of incompressible fluids, this reduces to
\begin{equation}
  \frac{\partial \Phi}{\partial t}+\bu\cdot\nabla\Phi + \nabla\cdot\bJ_{_\Phi}=0.
\end{equation}
At the boundary $\partial\Omega_1(t)=\Sigma_1(t)\bigcup\Sigma(t)$, we impose the following boundary conditions for the flux,
\begin{subequations}
\label{eqn:sflux12}
\begin{align}
\label{eqn:sflux1}
  \bn_w\cdot\bJ_{_\Phi}&=0,\qquad\mbox{on}\quad{\Sigma_1(t)},\\
  \bn\cdot\bJ_{_\Phi}&=S,\qquad\mbox{on}\quad{\Sigma(t)},
  \label{eqn:sflux2}
\end{align}
\end{subequations}
where $S$ is the surface-generating adsorption source for the interfacial surfactant concentration.

Similarly, the conservation law for the mass of the surfactants on the interface $\Sigma$ states that for any surface parcel $\hat{\Sigma}\subset\Sigma$,
\begin{align}
\label{eqn:scon}
  \frac{\mathrm{d}}{\dt}\int_{\hat{\Sigma}}\Gamma\,\dH^{d-1}+\int_{\partial\hat{\Sigma}}\bJ_{_\Gamma}\cdot\hat{\mathbf{m}}\,\dH^{d-2}=\int_{\hat{\Sigma}}S\,\dH^{d-1},
\end{align}
where $\bJ_{_\Gamma}$ is the surface diffusion flux tangential to the surface $\hat{\Sigma}$ and $\hat{\mathbf{m}}$ is the outward conormal of $\hat{\Sigma}$ at $\partial\hat{\Sigma}$. Again, applying the transport formula in~\eqref{eqn:sderivative} and using the surface divergence theorem in \eqref{eqn:integrationbyparts1}, the left hand side of \eqref{eqn:scon} can be recast as
\begin{align*}
  &\int_{\hat{\Sigma}}\left(\frac{\rD\Gamma}{\rD t} + \Gamma\,\nabla_s\cdot\vec u\right)\,\dH^{d-1}+\int_{\partial\hat{\Sigma}}\bJ_{\Gamma}\cdot\hat{\mathbf{m}}\,\dH^{d-2}\\
  &=\int_{\hat{\Sigma}}\left(\frac{\rD\Gamma}{\rD t}+\Gamma\,\nabla_s\cdot\vec u\right)\,\dH^{d-1} +\int_{\hat{\Sigma}}\nabla_s\cdot\bJ_{_\Gamma}\,\dH^{d-1}-\int_{\hat{\Sigma}}\kappa\,\bn\cdot\bJ_{_\Gamma}\dH^{d-1},
\end{align*}
 where we have introduced the surface gradient operator $\nabla_s$ whose definition is given in \ref{sec:appA}. Since $\bn\cdot\bJ_{_\Gamma}=0$ and $\hat{\Sigma}$ is arbitrary, we obtain the differential form of the conservation law on the surface $\Sigma$,
\begin{equation}
\label{eqn:ssdynamic}
  \frac{\rD \Gamma}{\rD t}+\Gamma\,\nabla_s\cdot\bu=-\nabla_s\cdot\bJ_\Gamma+S.
\end{equation}
At the contact line $\Lambda$, zero flux  implies that
\begin{align}
\label{eqn:sflux3}
\mathbf{m}\cdot\bJ_{_\Gamma}=0.
\end{align}
In comparison with the dynamic equation
of the bulk surfactant concentration in \eqref{eqn:bsdynamic}, there is an additional term  $\Gamma\,\nabla_s\cdot\vec u$ and a source term $S$ that comes from the bulk flux. Since the surface
can be either stretched or contracted, in general there is no divergence free condition on the surface, i.e., $\nabla_s\cdot\vec u\neq 0$.

The derivation of \eqref{eqn:bsdynamic} and \eqref{eqn:ssdynamic} reveals the transport properties of the surfactants in the bulk and on the interface. This is utilized in the weak formulation presented in \cref{sec:dmodel}.

\subsection{Dissipation for the total free energy}
We consider the dissipation rate of the total free energy defined in \eqref{eq:Tenergy}. We assume the usual non-penetration condition for the velocity on the solid wall
\begin{align}
\label{eqn:nwvelocity}
\vec u\cdot\vec n_w = 0, \quad{\rm on}\quad \Sigma_1\cup\Sigma_2.
\end{align}
 For the fluid kinetic energy, by using the dynamic equations \eqref{eqn:fluiddynamic}, the divergence theorem in the tensor form and the condition \eqref{eqn:nwvelocity}, we obtain
\begin{align}
\frac{{\rm d}}{{\rm d}t}&\left(\sum_{i=1,2}\int_{\Omega_i(t)}\frac{1}{2}\rho_i|\vec u|^2\;\dH^d\right)\nonumber\\
&=\sum_{i=1,2}\int_{\Omega_i}\rho_i\vec u\cdot\left(\frac{\partial\vec u}{\partial t} + \vec u\cdot\nabla\vec u\right)\;\dH^d\nonumber\\
&=-\sum_{i=1,2}\int_{\Omega_i}2\eta_i\norm{D(\bu)}_F^2\;\dH^d  + \sum_{i=1,2}\int_{\Sigma_i}\mathcal{P}_{w}(\sigma_i\,\vec n_w)\cdot \bu_s\,\dH^{d-1}+\int_{\Sigma}\vec u\cdot[p\mathbf{I}-\sigma_i]_1^2\cdot\vec n\,\dH^{d-1},\label{eq:dissipation1}
\end{align}
where $[\cdot]_1^2$ denotes the jump value from fluid 1 to fluid 2, $\mathcal{P}_{w}:=\left(\mathbf{I} - \vec n_w\otimes\vec n_w\right)$ is the projection operator on the solid surface, $\vec u_s=\mathcal{P}_w\bu$ is the slip velocity of fluids on the solid wall, and $\norm{\cdot}_F$ denotes the Frobenius norm.

Next we take the time derivative of the bulk chemical energy. Making use of the dynamic equation of $\Phi$ in \eqref{eqn:bsdynamic}, the transport formula in \eqref{eqn:dderivative}, the divergence theorem and the divergence free condition \eqref{eq:fluiddynamic2}, we have
\begin{align}
  \frac{\mathrm{d}}{\dt}\int_{\Omega_1(t)}f(\Phi)\dH^d&=\int_{\Omega_1}\left[f'(\Phi)\left(\frac{\partial\Phi}{\partial t}+\vec u\cdot\nabla\Phi\right) + f(\Phi)\,\nabla\cdot\vec u)\right]\dH^d \nonumber\\
  &=\int_{\Omega_1}\left[-f'(\Phi)\nabla\cdot\bJ_{\Phi} + (f(\Phi)-f'(\Phi)\Phi)\,\nabla\cdot\vec u)\right]\dH^d \nonumber\\
  &=-\int_{\Sigma}f'(\Phi)\,\bn\cdot\bJ_{_\Phi}\,\dH^{d-1}-\int_{\Sigma_1}f'(\Phi)\,\bn_w\cdot\bJ_{_\Phi}\dH^{d-1}+\int_{\Omega_1}\nabla f'(\Phi)\cdot\bJ_{_\Phi}\dH^d\nonumber\\
  &=-\int_{\Sigma}f'(\Phi)\,S\,\dH^{d-1}+\int_{\Omega_1}\nabla f'(\Phi)\cdot\bJ_{_\Phi}\dH^d,
  \label{eq:dissipation2}
\end{align}
where we have used the boundary conditions in \eqref{eqn:sflux12}. We remark that the surfactant induced osmotic pressure term $f(\Phi)-f'(\Phi)\Phi$ plays no role due to the incompressibility condition. However, the osmotic pressure is important in the case of compressible flows.

Similarly, applying the time derivative to the interfacial energy of the fluid interface and using the transport formula in \eqref{eqn:sderivative} as well as the dynamic equation for the interfacial surfactants in \eqref{eqn:ssdynamic}, we have
\begin{align}
\label{eq:dss3}
\frac{\mathrm{d}}{\dt}\int_{\Sigma}g(\Gamma)\,\dH^{d-1} &= \int_{\Sigma}\left[g^\prime(\Gamma)\,\frac{\rD\Gamma}{\rD t} +g(\Gamma)\,\nabla_s\cdot\vec u\right]\,\dH^{d-1}\\
&=\int_{\Sigma}\left(g(\Gamma) - g'(\Gamma)\,\Gamma\right)\,\nabla_s\cdot\vec u\,\dH^{d-1} - \int_{\Sigma}g'(\Gamma)\,\nabla_s\cdot\bJ_{_\Gamma}\,\dH^{d-1} + \int_{\Sigma}g'(\Gamma)\,S\,\dH^{d-1}.\nn
\end{align}
The surface tension of $\Sigma$ is defined as the Legendre transform of $g(\Gamma)$:
\begin{align}
\label{eq:gammaGamma}
\gamma(\Gamma):=g(\Gamma)-g'(\Gamma)\,\Gamma.
\end{align}
Using integration by parts formula in \eqref{eqn:integrationbyparts2} and the no flux boundary condition \eqref{eqn:sflux3},  Eq.~\eqref{eq:dss3} can be recast as:
\begin{align}
 &\frac{\mathrm{d}}{\dt}\int_{\Sigma}g(\Gamma)\,\dH^{d-1}\nonumber\\
  &\quad=\int_{\Lambda}\gamma(\Gamma)\,\bu\cdot\mathbf{m}\,\dH^{d-2}+\int_{\Sigma}\Bigl\{\bu\cdot\left(\gamma(\Gamma)\,\kappa\,\bn - \nabla_s\gamma(\Gamma)\right)+\bJ_{_\Gamma}\cdot\nabla_sg'(\Gamma)+g'(\Gamma)\,S\Bigr\}\,\dH^{d-1},
\label{eq:dissipation3}
\end{align}
where $\kappa=\nabla_s\cdot\bn$ is the mean curvature, and we have used the orthogonality $\bJ_{_\Gamma}\cdot\vec n=0$ and the zero flux condition \eqref{eqn:sflux3}.

The time derivative of the surface free energy at wall is simply given by
\begin{equation}
  \frac{\mathrm{d}}{\dt}\int_{\Sigma_1}(\gamma_1-\gamma_2)\dH^{d-1}=\int_{\Lambda}(\gamma_1-\gamma_2)\,\bu\cdot\mathbf{m}_1\,\dH^{d-2}.
  \label{eq:dissipation4}
\end{equation}
%

Collecting results in Eqs.~\eqref{eq:dissipation1}-\eqref{eq:dissipation4}, we arrive at
\begin{align}
  \frac{\mathrm{d}}{\dt}W(t)=&-\sum_{i=1}^2\int_{\Omega_i}2\eta_i\,\norm{D(\bu)}_F^2\,\dH^d+\int_{\Omega_1}\nabla f'(\Phi)\cdot\bJ_{_\Phi}\,\dH^d+\sum_{i=1}^2\int_{\Sigma_i}\bu_s\cdot\mathcal{P}_w(\sigma_i\,\bn_w)\,\dH^{d-1}\nonumber\\
  &+\int_{\Sigma}\Bigl\{\bu\cdot\left[p\mathbf{I}-\sigma_i\right]_1^2\cdot\bn +\bu\cdot\Bigl(\gamma(\Gamma)\,\kappa\,\bn-\nabla_s\gamma(\Gamma)\Bigr)\Bigr\}\,\dH^{d-1}+\int_{\Sigma}\bJ_{_\Gamma}\cdot\nabla_sg'(\Gamma)\dH^{d-1}\nonumber\\
  &+\int_{\Sigma}\Bigl(g'(\Gamma)-f'(\Phi)\Bigr)\,S\,\dH^{d-1}+\int_{\Lambda}\Bigl(\gamma(\Gamma)\,\cos\theta_d-(\gamma_2-\gamma_1)\Bigr)\,u_{l}\,\dH^{d-2},
  \label{eqn:energyderivative}
\end{align}
where $u_{l}=\bu\cdot\mathbf{m}_1$ is the normal velocity of the contact line $\Lambda$, and $\theta_d$ is the dynamic contact angle between $\mathbf{m}$ and $\mathbf{m}_1$.

Based on second law of thermodynamics, we choose the constitutive relations so that the total free energy has non-positive dissipation. Here we require that each term on the right hand side of \eqref{eqn:energyderivative} is non-positive. This implies:
\begin{enumerate}[label=(\roman*)]
\item  The dissipation induced by the bulk and surface diffusion gives the constraints
\begin{subequations}
\label{eqn:fgconstraints}
\begin{align}
\label{eq:ffconstraints}
f^{\prime\prime}(\Phi)\,\bJ_{_\Phi}\cdot\nabla\Phi\leqslant0,\quad{\rm in}\quad\Omega_1(t)\\
g^{\prime\prime}(\Gamma)\,\bJ_{_\Gamma}\cdot\nabla_s\Gamma\leqslant0,\quad{\rm on}\quad \Sigma(t).
\label{eq:ggconstraints}
\end{align}
\end{subequations}
\item  The dissipations on the solid wall lead to the slip boundary condition for the velocity
\begin{equation}
\label{eq:slipvelocity}
  \mathcal{P}_w(\sigma_i\,\bn_w)=\mathbf{f}_i^w(\bu_{s}),\quad{\rm on}\quad \Sigma_i(t)\,(i=1,2),
\end{equation}
where the map $\mathbf{f}_i^w$ satisfies
\begin{align}
\label{eq:slipcontraints}
\mathbf{f}_i^w(\bu_{s})\cdot\bu_{s}\leqslant 0.
\end{align}
\item  On the fluid interface, the energy dissipation of the surfactant adsorption and desorption process requires the constraint
\begin{align}
\label{eq:Sourceconstraints}
(g^\prime(\Gamma) - f^\prime(\Phi))\,S\leqslant0,\quad{\rm on}\quad \Sigma(t).
\end{align}
\item  On the fluid interface, there is no interfacial dissipation. This gives the interfacial jump condition
\begin{equation}
\label{eq:LYequation}
  [p\mathbf{I}-\sigma_i]_1^2\cdot\bn=-\gamma(\Gamma)\,\kappa\,\bn+\nabla_s\gamma(\Gamma),\quad{\rm on}\quad \Sigma(t).
\end{equation}
This is the Laplace-Young interface condition describing the balance of the stress jump of the fluids and the surface force. The force acting on the fluid interface has two components: the capillary force $\gamma(\Gamma)\,\kappa\,\vec n$ in the normal direction and the Marangoni force $\nabla_s\gamma(\Gamma)$ in the tangential direction.
\item The dissipation at the contact line gives the condition for the dynamic contact angle
\begin{equation}
\label{eq:CLvelocity}
  \gamma(\Gamma)\cos\theta_d-(\gamma_2-\gamma_1)=f_{CL}(u_l),\quad{\rm at}\quad \Lambda(t),
\end{equation}
where the function $f_{CL}$ satisfies
\begin{align}
\label{eq:CLconstraints}
f_{CL}(u_l)u_l\leqslant0.
\end{align}
\end{enumerate}
 In summary, the dynamic equations \eqref{eqn:fluiddynamic}, \eqref{eqn:bsdynamic} and \eqref{eqn:ssdynamic} together with the no flux boundary conditions for the surfactants in \eqref{eqn:sflux12} and \eqref{eqn:sflux3}, the interface conditions in \eqref{eq:LYequation}, the usual non-penetration condition in \eqref{eqn:nwvelocity}, the condition for the slip velocity in \eqref{eq:slipvelocity} as well as the contact angle condition in \eqref{eq:CLvelocity} form a model for the moving contact lines with surfactants.

\subsection{Constitutive relations and adsorption isotherms}
 In the following, we will discuss the typical choices of $f_i^w$, $f_{CL}$, $f$, $g$, the surfactant flux $\bJ_{_\Phi}$, $\bJ_{_\Gamma}$ and the source term $S$ such that the constraints in \eqref{eq:slipcontraints}, \eqref{eq:CLconstraints}, \eqref{eqn:fgconstraints} and \eqref{eq:Sourceconstraints} can be well satisfied. When the slip velocity is not too large, we approximate $f_i^w$ and $f_{CL}$ using linear response:
\begin{align*}
f_i^w(\bu_s) = -\beta_i\bu_s,\qquad f_{CL}(u_l) = -\beta^{*} u_l,
\end{align*}
where $\beta_i$ and $\beta^{*}$ are the friction coefficients on the solid substrate and at the contact line, respectively. This gives the Naiver slip condition
\begin{align}
\mathcal{P}_w(\sigma_i\bn_w) = - \beta_i\,\bu_s,
\label{eq:NVslip}
\end{align}
and the contact angle condition
\begin{align}
\label{eq:DYangle}
\gamma(\Gamma)\cos\theta_d-(\gamma_2-\gamma_1)=-\beta^{*}u_l.
\end{align}
Eq. \eqref{eq:DYangle} states that the unbalanced Young's force is balanced by the friction force at the contact line.
For the bulk and surface flux $\bJ_{_\Phi},~\bJ_\Gamma$ as well as the adsorption source term $S$, we consider two special cases:
\begin{itemize}
\item [i)] Langmuir isotherm: we assume that the bulk surfactants are dilute so that $\Phi\ll1$. Then we can use the dilute approximation of $f(\Phi)$:
\begin{align*}
  f(\Phi)= \mu_b\Phi+RT\Phi_{\infty}(\frac{\Phi}{\Phi_{\infty}}\ln\frac{\Phi}{\Phi_{\infty}}-\frac{\Phi}{\Phi_{\infty}}),
\end{align*}
where $\mu_b$ is the standard chemical potential in the bulk, $R=k_BN_a$ is the gas constant with $k_B$ and $N_a$ being Boltzmann constant and Avgadro constant respectively, $T$ is the absolute temperature, and $\Phi_\infty$ is the maximum volumetric mole of the surfactant.
It is easy to calculate that
$f'(\Phi)=\mu_b+RT\ln\frac{\Phi}{\Phi_{\infty}}$ and $f''(\Phi)=\frac{RT}{\Phi}>0$. Applying linear response to \eqref{eq:ffconstraints}, we recover the Fick's law for the diffusion flux
\begin{align}
\label{eq:DPhi}
\bJ_{_\Phi}=-D_{_\Phi}\nabla \Phi,\quad{\rm with}\quad D_{_\Phi}>0,
\end{align}
where $D_{_\Phi}$ is the diffusion coefficient for the bulk surfactants.

The surface free energy follows from the entropy of mixing,
\begin{equation*}
  g(\Gamma)=\gamma_0+\mu_s\Gamma+RT\Gamma_\infty\Big(\frac{\Gamma}{\Gamma_\infty}\ln\frac{\Gamma}{\Gamma_\infty}+(1-\frac{\Gamma}{\Gamma_\infty})\ln(1-\frac{\Gamma}{\Gamma_\infty})\Big),
\end{equation*}
where $\gamma_0$ is the surface tension coefficient of a clean fluid interface, $\mu_s$ is the standard chemical potential on the surface, and $\Gamma_{\infty}$ is the interfacial surfactant concentration at the maximum packing. Its derivative is calculated as $g'(\Gamma)=\mu_s+RT\ln\frac{\Gamma}{\Gamma_\infty-\Gamma}$ and $g''(\Gamma)=\frac{RT\Gamma_\infty}{\Gamma(\Gamma_\infty-\Gamma)}>0$. Based on \eqref{eq:gammaGamma}, the interfacial tension is then given by
\begin{align}
\gamma(\Gamma)=\gamma_0+RT\Gamma_\infty\ln(1-\frac{\Gamma}{\Gamma_\infty}),\nn
\end{align}
which is Langmuir equation of state. Applying linear response to \eqref{eq:ggconstraints} gives the Fick's law for the surface diffusion flux
\begin{align}
\label{eq:DGamma}
\bJ_{_\Gamma}=-D_{_\Gamma}\nabla_s\Gamma,\quad{\rm with}\quad D_{_\Gamma}>0,
\end{align}
where $D_{_\Gamma}$ is the diffusion coefficient for the interfacial surfactants.

At equilibrium, the balance in the chemical potential $f'(\Phi^{eq})=g'(\Gamma^{eq})$ gives rise to
\begin{equation}
\frac{\Gamma^{eq}/\Gamma_\infty}{(1-\Gamma^{eq}/\Gamma_\infty)\Phi^{eq}/\Phi_\infty}=\lambda,\nn
\end{equation}
where $\lambda=\exp\Big(\frac{\mu_b-\mu_s}{RT}\Big)$ is the equilibrium constant. Away from equilibrium, the thermodynamic constraint in \eqref{eq:Sourceconstraints} implies $SRT\ln\frac{\Gamma/\Gamma_\infty}{\lambda(1-\Gamma/\Gamma_\infty)\Phi/\Phi_\infty}\leqslant0.$ By the simple inequality $(a-b)\ln(\frac{a}{b})\geqslant0$ for any $a>0$ and $b>0$, we can choose $a=\lambda(1-\Gamma/\Gamma_\infty)\Phi/\Phi_\infty$ and $b=\Gamma/\Gamma_\infty$ so that the thermodynamically consistent adsorption rate is given by
\begin{align}
S= k_{ad}\frac{\Phi}{\Phi_\infty}(1-\frac{\Gamma}{\Gamma_\infty})-k_d\frac{\Gamma}{\Gamma_\infty},\nn
\end{align}
where $k_{ad}=\lambda k_d$ and $k_d$ are the adsorption and desorption coefficients respectively. This gives the Langmuir adsorption isotherm. We also refer to an early work of thin film model with soluble surfactants, in which the isotherm is derived based on the linear response near the equilibrium \cite{thiele16gradient}.

The Langmuir adsorption isotherm can also be understood using the stoichiometric theory of a reversible chemical reaction
\begin{equation*}
  A+B\rightleftharpoons AB,
\end{equation*}
where $A$ represents an empty adsorption site, $B$ represents free surfactant ion nearby the interface, and $AB$ stands for an adsorbed surfactant ion. The law of mass action gives the reaction rate
\begin{equation*}
  S(A,B,AB)=k_{ad}c_{A}c_{B}-k_dc_{AB},
\end{equation*}
where $c_{A}=1-\frac{\Gamma}{\Gamma_\infty}$ is the concentration of empty adsorption sites, $c_B=\frac{\Phi}{\Phi_\infty}$ is the concentration of free surfactant ions nearby the interface, and $c_{AB}=\frac{\Gamma}{\Gamma_\infty}$ is the concentration of adsorbed surfactant ions. The connection between the law of mass action and the variational principle was discussed in detail in \cite{wang20field}.
\item [ii)] Frumkin isotherm: we modify the surface free energy density by adding one more term concerning the interaction between surfactant molecules on the interface,
\begin{equation*}
  g(\Gamma)=\gamma_0+\mu_s\Gamma+RT\,\Gamma_\infty\Big(\frac{\Gamma}{\Gamma_\infty}\ln\frac{\Gamma}{\Gamma_\infty}+(1-\frac{\Gamma}{\Gamma_\infty})\ln(1-\frac{\Gamma}{\Gamma_\infty})\Big)-\frac{K}2\Big(\frac{\Gamma}{\Gamma_\infty}\Big)^2,
\end{equation*}
where $K$ is the interaction constant. Its derivative can be calculated accordingly, $g'(\Gamma)=\mu_s+RT\ln\frac{\Gamma}{\Gamma_\infty-\Gamma}-A\frac{\Gamma}{\Gamma_\infty^2}$ and $g''(\Gamma)=\frac{RT\Gamma_\infty^3-K\Gamma(\Gamma_\infty-\Gamma)}{\Gamma_\infty^2\Gamma(\Gamma_\infty-\Gamma)}$. In order to apply the Fick's flux $\bJ_{_\Gamma}=-D_{_\Gamma}\nabla_s\Gamma$ in \eqref{eq:ggconstraints}, it is required that $RT\Gamma_\infty^3-K\Gamma(\Gamma_\infty-\Gamma)\geqslant0$ for any $\Gamma$, which holds if $K\leqslant4RT\Gamma_\infty$. The interfacial tension is
\begin{align}
\gamma(\Gamma)=\gamma_0+RT\Gamma_\infty\ln(1-\frac{\Gamma}{\Gamma_\infty})+\frac{K}2\Big(\frac{\Gamma}{\Gamma_\infty}\Big)^2,\nn
\end{align}
which is Frumkin equation of state. The same argument gives rise to the Frumkin adsorption rate
\begin{align}
S= k_{ad}\frac{\Phi}{\Phi_\infty}(1-\frac{\Gamma}{\Gamma_\infty})-k_d\frac{\Gamma}{\Gamma_\infty}e^{-K\Gamma/RT\Gamma_\infty^2}.\nn
\end{align}
\end{itemize}
With different choices of $g(\Gamma)$, we could recover different isotherms. We summarize them in Table \ref{table:isotherm}. These are consistent with the results in \cite{Danov02adsorption, kralchevsky16chem, Garcke2014diffuse}.
\begin{table}
  \begin{tabular}{|c|c|c|c|}
    \hline
    Isotherm & \tabincell{c}{Surface energy\\ density $g(\Gamma)$} & \tabincell{c}{Equation of\\ state $\gamma(\Gamma)$} & \tabincell{c}{Adsorption\\ rate $S(\Phi,\Gamma)$} \\
    \hline
    Henry & \tabincell{c}{$\gamma_0+\mu_s\Gamma+$\\$RT\Gamma_\infty(\frac{\Gamma}{\Gamma_\infty}\ln\frac{\Gamma}{\Gamma_\infty}-\frac{\Gamma}{\Gamma_\infty})$} & $\gamma_0-RT\Gamma$ & $k_{ad}\frac{\Phi}{\Phi_\infty}-k_d\frac{\Gamma}{\Gamma_\infty}$ \\
    \hline
    Freundlich & \tabincell{c}{$\gamma_0+\mu_s\Gamma+$\\$\frac{RT\Gamma_\infty}{m}(\frac{\Gamma}{\Gamma_\infty}\ln\frac{\Gamma}{\Gamma_\infty}-\frac{\Gamma}{\Gamma_\infty})$} & $\gamma_0-\frac1mRT\Gamma$ & \tabincell{c}{$k_{ad}\lambda^{m-1}\Big(\frac{\Phi}{\Phi_\infty}\Big)^m$\\$-k_d\frac{\Gamma}{\Gamma_\infty}$} \\
    \hline
    Langmuir & \tabincell{c}{$\gamma_0+\mu_s\Gamma+$\\$RT\Gamma_\infty\Big(\frac{\Gamma}{\Gamma_\infty}\ln\frac{\Gamma}{\Gamma_\infty}+$\\$(1-\frac{\Gamma}{\Gamma_\infty})\ln(1-\frac{\Gamma}{\Gamma_\infty})\Big)$} & $\gamma_0+RT\Gamma_\infty\ln(1-\frac{\Gamma}{\Gamma_\infty})$ & \tabincell{c}{$k_{ad}\frac{\Phi}{\Phi_\infty}(1-\frac{\Gamma}{\Gamma_\infty})$\\$-k_d\frac{\Gamma}{\Gamma_\infty}$} \\
    \hline
    Volmer & \tabincell{c}{$\gamma_0+\mu_s\Gamma+$\\$RT\Gamma_\infty\Big(\frac{\Gamma}{\Gamma_\infty}\ln\frac{\Gamma}{\Gamma_\infty}-\frac{\Gamma}{\Gamma_\infty}$\\$-\frac{\Gamma}{\Gamma_\infty}\ln(1-\frac{\Gamma}{\Gamma_\infty})\Big)$}  & $\gamma_0-RT\Gamma_\infty\Big(\frac{\Gamma}{\Gamma_\infty}+\frac{(\frac{\Gamma}{\Gamma_\infty})^2}{1-\frac{\Gamma}{\Gamma_\infty}}\Big)$ & \tabincell{c}{$k_{ad}\frac{\Phi}{\Phi_\infty}-$\\$k_d\frac{\Gamma}{\Gamma_\infty-\Gamma}\exp(\frac{\Gamma}{\Gamma_\infty-\Gamma})$} \\
    \hline
    Frumkin & \tabincell{c}{$\gamma_0+\mu_s\Gamma+$\\$RT\Gamma_\infty\Big(\frac{\Gamma}{\Gamma_\infty}\ln\frac{\Gamma}{\Gamma_\infty}+$\\$(1-\frac{\Gamma}{\Gamma_\infty})\ln(1-\frac{\Gamma}{\Gamma_\infty})\Big)$\\$-\frac{K}2\Big(\frac{\Gamma}{\Gamma_\infty}\Big)^2$}  & \tabincell{c}{$\gamma_0+RT\Gamma_\infty\ln(1-\frac{\Gamma}{\Gamma_\infty})$\\$+\frac{K}2\Big(\frac{\Gamma}{\Gamma_\infty}\Big)^2$} & \tabincell{c}{$k_{ad}\frac{\Phi}{\Phi_\infty}(1-\frac{\Gamma}{\Gamma_\infty})$\\$-k_d\frac{\Gamma}{\Gamma_\infty}\exp(-\frac{K\Gamma}{RT\Gamma_\infty^2})$} \\
    \hline
    van de Waals & \tabincell{c}{$\gamma_0+\mu_s\Gamma+$\\$RT\Gamma_\infty\Big(\frac{\Gamma}{\Gamma_\infty}\ln\frac{\Gamma}{\Gamma_\infty}-\frac{\Gamma}{\Gamma_\infty}$\\$-\frac{\Gamma}{\Gamma_\infty}\ln(1-\frac{\Gamma}{\Gamma_\infty})\Big)$\\$-\frac{K}2\Big(\frac{\Gamma}{\Gamma_\infty}\Big)^2$}   & \tabincell{c}{$\gamma_0-RT\Gamma_\infty\Big(\frac{\Gamma}{\Gamma_\infty}+\frac{(\frac{\Gamma}{\Gamma_\infty})^2}{1-\frac{\Gamma}{\Gamma_\infty}}\Big)$\\$+\frac{K}2\Big(\frac{\Gamma}{\Gamma_\infty}\Big)^2$} & \tabincell{c}{$k_{ad}\frac{\Phi}{\Phi_\infty}-$\\$k_d\frac{\Gamma}{\Gamma_\infty-\Gamma}\exp(\frac{\Gamma}{\Gamma_\infty-\Gamma}$\\$-\frac{K\Gamma}{RT\Gamma_\infty^2})$} \\
    \hline
  \end{tabular}
  \caption{Different kinetic models for the adsorption isotherm. For all models presented here, the bulk energy density is of the same form as $f(\Phi)= \mu_b\Phi+RT\Phi_\infty(\frac{\Phi}{\Phi_\infty}\ln\frac{\Phi}{\Phi_\infty}-\frac{\Phi}{\Phi_\infty})$. The parameter $m$ measures the adsorption intensity in the Freundlich model. Except for the Freundlich model, all isotherms reduce to the Henry isotherm as $\Gamma\rightarrow0$.}
  \label{table:isotherm}
\end{table}

Using the interface condition in \eqref{eq:LYequation}, the boundary conditions in \eqref{eq:NVslip}, \eqref{eq:DYangle}, and the flux \eqref{eq:DPhi} and \eqref{eq:DGamma}, the energy law in \eqref{eqn:energyderivative} becomes
\begin{align}
\frac{{\rm d}}{{\rm d}t}W(t)&=-\sum_{i=1,2}\int_{\Omega}2\eta_i\,\norm{D(\vec u)}_F^2\,\dH^d - \int_{\Omega_1}D_{_\Phi}\,f^{\prime\prime}(\Phi)|\nabla\Phi|^2\,\dH^{d} - \sum_{i=1,2}\int_{\Sigma_i}\beta_i|\bu_s|^2\,\dH^{d-1} \nonumber\\
&\quad-  \int_{\Sigma}D_{_\Gamma}\,g^{\prime\prime}(\Gamma)\,|\nabla_s\Gamma|^2\,\dH^{d-1} + \int_{\Sigma}\Bigl[g'(\Gamma) - f'(\Phi)\Bigr]\,S\,\dH^{d-1} -\int_{\Lambda} \beta^*\,u_l^2\,\dH^{d-2}\leqslant0,
\end{align}
where for the last inequality we have used the fact that
\begin{align*}
f^{\prime\prime}(\Phi)>0,\quad g^{\prime\prime}(\Gamma)>0,\qquad [g'(\Gamma) - f'(\Phi)]\,S\leqslant0.
\end{align*}
\section{The dimensionless model in two spatial dimensions and its weak form}
\label{sec:dmodel}

\begin{figure}[!htp]
\centering
\includegraphics[width=0.75\textwidth]{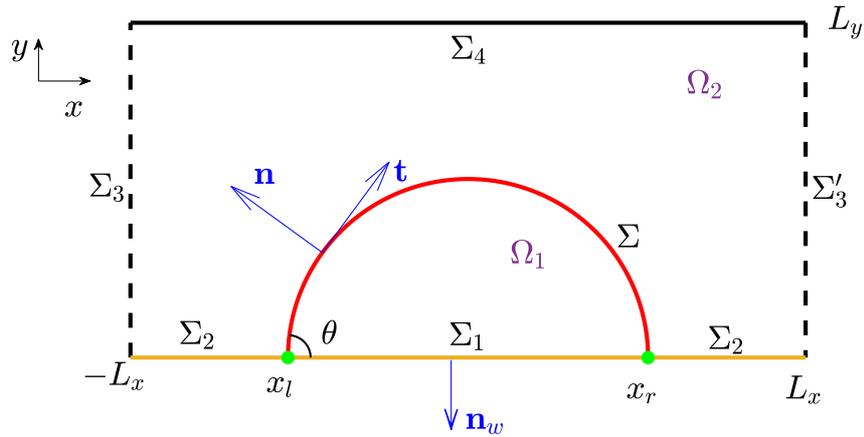}
\caption{Moving contact lines with soluble surfactants in a bounded domain $\Omega=\Omega_1\cup\Omega_2=[-L_x,~L_x]\times[0,~L_y]$ in two spatial dimensions, where $x_l,x_r$ corresponds to the left and right contact points, respectively. }
\label{fig:model}
\end{figure}
For the purpose of numerical implementation,  we consider the dynamic system with simplified Langmuir isotherm ($\mu_b=\mu_s=0$) in a bounded domain in the 2d space. As shown in Fig.~\ref{fig:model}, we assume the periodic structure along the horizontal direction and use Cartesian coordinates, where the substrate is on the $x$-axis. We define: $\vec t_w=(1,0)^T$ and $\bn_w=(0,-1)^T$ as the unit tangential and normal vectors of the liquid-solid interface, $\vec n$ and $\vec t$ as the unit tangential and normal vectors of the fluid-fluid interface, and $x_l$ and $x_r$  as the left and right contact points, respectively.

\subsection{The dimensionless model}
We write the governing equations and boundary/interface conditions in their dimensionless form. By choosing $L$ and $U$ as the characteristic length and velocity, respectively, we rescale the physical quantities
\begin{equation*}
\hat{\rho}_i = \frac{\rho_i}{\rho_1},\quad \hat{\eta}_i = \frac{\eta_i}{\eta_1},
\quad\hat{\beta}_i = \frac{\beta_i}{\beta_1}, \quad \hat{\beta}^* = \frac{\beta^*}{\eta_1},
\quad\hat{\gamma}_i = \frac{\gamma_i}{\gamma_0},\quad \hat{\vec x} = \frac{\vec x}{L},
\end{equation*}
\begin{equation*}
\quad \hat{\vec u} = \frac{\vec u}{U},\quad\hat{t}
= \frac{Ut}{L},\quad \hat{p} = \frac{p}{\rho_1 U^2},\quad \hat{\kappa} = L\kappa,\quad \hat{\Phi} = \frac{\Phi}{\Phi_{\infty}},\quad \hat{\Gamma} = \frac{\Gamma}{\Gamma_{\infty}}.
\end{equation*}
Then the Reynolds number $Re$, the Capillary number $Ca$, the slip length $l_s$
and the Weber number $We$ are given by
\begin{equation*}
Re = \frac{\rho_1UL}{\eta_1},\quad Ca = \frac{\eta_1U}{\gamma_0},
\quad l_s = \frac{\eta_1}{\beta_1L},\quad We =\frac{\rho_1\,U^2L}{\gamma_0} = Re\cdot Ca.
\end{equation*}
Besides, the Biot number $Bi$, the surfactant elasticity $E$, the adsorption depth $Da$, the bulk P\'{e}clet number $Pe_{_\Phi}$ and the interface P\'{e}clet number $Pe_{_\Gamma}$ are defined as
\begin{equation*}
Bi = \frac{k_d L}{U\Gamma_\infty},\quad E = \frac{RT\Gamma_{\infty}}{\gamma_0},\quad Da = \frac{\Gamma_{\infty}}{L \Phi_{\infty}},\quad Pe_{_\Phi} = \frac{U\,L}{D_{_\Phi}},\quad Pe_{_\Gamma}=\frac{U\,L}{D_{_\Gamma}}.
\end{equation*}

In the following, we will drop the hats on the dimensionless variables for ease of presentation. In the dimensionless form, the governing equations for the fluid dynamics in $\Omega_i(t)(i=1,2)$ read:
\begin{subequations} \label{eq:model12}
\begin{numcases}{}
\label{eqn:model1}
\rho_i\,\left(\frac{\partial\vec u}{\partial t} + \vec u\cdot\nabla\vec u\right) +\nabla\cdot\mathbf{T} =\vec 0,\\
\label{eqn:model2}
\nabla\cdot\vec u = 0,
\end{numcases}
\end{subequations}
where $\mathbf{T} = p\mathbf{I} - \frac{1}{Re}\sigma_i$, and $\sigma_i=\eta_i (\nabla\vec u+(\nabla\vec u)^T)$ is the dimensionless viscous stress tensor. The governing equations are subject to the following boundary/interface conditions:
\begin{itemize}
\item [(i)] The interface conditions on $\Sigma(t)$:
\begin{subequations} \label{eq:bd1}
\begin{align}
\label{eq:k}
  &\bigl[\vec u\bigr]^2_1 = 0,\qquad We\,\bigl[\mathbf{T}\bigr]_1^2\cdot\vec n = -\gamma(\Gamma) \kappa\,\vec n + \nabla_{s}\gamma(\Gamma),\\
&v_n = \vec u|_{_{\Sigma(t)}}\cdot\vec n.\qquad
\kappa = -\partial_{ss}\vec X\cdot\vec n,
\label{eq:cv}
\end{align}
\end{subequations}
where $v_n$ is the interface normal velocity, $\vec X$ denotes the fluid interface, $\partial_{ss}$ represents the second order partial derivative with respect to the arclength parameter, and the dimensionless surface tension (rescaled by $\gamma_0$) is given by
\begin{align}
\gamma(\Gamma) = 1+ E\ln(1 - \Gamma).
\end{align}
\item [(ii)] The boundary conditions on $\Sigma_1(t)\cup\Sigma_2(t)$:
%
\begin{align}
\label{eqn:bd2}
\vec u\cdot\vec n_w=0,\qquad
l_s(\bt_w\cdot\sigma_i\cdot\vec n_w) =  -\beta_i u_s.
\end{align}
%
%
\item [(iii)] The condition for the dynamic contact angles:
%
\begin{align}\label{eqn:bd3}
& \frac{1}{Ca}\left(\gamma(\Gamma_l)\cos\theta_d^l - \cos\theta_Y\right)=\beta^*\dot{x}_l(t), \quad
& \frac{1}{Ca}\left(\gamma(\Gamma_r)\cos\theta_d^r - \cos\theta_Y\right)=-\beta^*\dot{x}_r(t),
\end{align}
%
where $\Gamma_{l,r}:=\Gamma|_{x=x_{l,r}}$, and $\theta_Y=\arccos(\frac{\gamma_2-\gamma_1}{\gamma_0})$ is the Young's angle for the clean fluid interface.
\item [(iv)] Periodic boundary conditions on $\Sigma_3\cup\Sigma_3^\prime$:
%
\begin{align}
\label{eqn:bd5}
\vec u(-L_x,y,t) = \vec u(L_x,y,t),\quad
\mathbf{T}(-L_x,y,t) = \mathbf{T}(L_x,y,t).
\end{align}
%
\item[(v)] The no-slip condition on the upper wall $\Sigma_4$:
\begin{equation}
\label{eqn:bd6}
\vec u = \vec 0.
\end{equation}
\end{itemize}

In dimensionless form, the governing equations for the surfactant concentrations read
\begin{subequations}
\label{eqn:phig}
\begin{numcases}{}
\label{eqn:phis}
\frac{\partial\Phi}{\partial t} + \vec u\cdot\nabla\Phi -\frac{1}{Pe_{_\Phi}}\nabla^2\Phi = 0,\qquad\vec x\in\Omega_1(t)\\
\frac{\rD\Gamma}{\rD t} + \Gamma\,\nabla_s\cdot\vec u - \frac{1}{Pe_{_\Gamma}}\nabla_s^2\Gamma = S(\Phi,\Gamma),\qquad\vec x\in\Sigma(t),
\label{eqn:gammas}
\end{numcases}
\end{subequations}
together with boundary conditions
\begin{subequations}
\label{eqn:surbbd}
\begin{align}
\label{eqn:surbd1}
\frac{1}{Pe_{_\Phi}}(\vec n\cdot\nabla\Phi) = -Da\, S(\Phi,\Gamma)&,\quad{\rm on}\;\Sigma(t),\\
\label{eqn:surbd2}
(\vec n_w\cdot\nabla\Phi) = 0&,\quad{\rm on}\;\Sigma_1(t).\\
(\vec t\cdot\nabla_s\Gamma) = 0&,\quad{\rm at}\;x_{l,r}(t),
\label{eqn:surbd3}
\end{align}
\end{subequations}
where the source term for Langmuir kinetic is given by
\begin{align}
S(\Phi,\Gamma) := Bi \left(\lambda\,\Phi(1- \Gamma) - \Gamma\right),\qquad 0\leqslant\Phi\leqslant 1, \quad 0\leqslant\Gamma\leqslant 1.
\end{align}

Equation \eqref{eq:model12} and \eqref{eqn:phig} together with the boundary\slash interface conditions \eqref{eq:bd1}-\eqref{eqn:bd6} and \eqref{eqn:surbbd} form a complete model for the dimensionless dynamic system in a bounded domain. In terms of the dimensionless variables, the dimensionless total energy of the system reads (rescaled by $\rho_1U^2L^2$)
\begin{equation}
\label{eqn:dimenenergy}
W(t)=\sum_{i=1,2}\int_{\Omega_i(t)}\frac{1}{2}\rho_i|\vec u|^2\;
\dH^2 + \frac{1}{We\cdot Da}\int_{\Omega_1(t)} f(\Phi)\,\dH^2 + \frac{1}{We} \int_{\Sigma(t)}g(\Gamma)\,\ds -\frac{\cos\theta_Y}{We}|\Sigma_1(t)|,
\end{equation}
where $|\Sigma_1(t)|$ is the total length of $\Sigma_1(t)$, $f$ and $g$ are given respectively as
\begin{align}
f(\Phi) = E\Phi\Bigl(\ln(\lambda\,\Phi) - 1\Bigr),\qquad g(\Gamma)= 1+ E\,\Bigl(\Gamma\ln\Gamma + (1 - \Gamma)\ln(1 - \Gamma)\Bigr).\nn
\end{align}
%

\begin{thm}
The dynamic system \eqref{eq:model12}--\eqref{eqn:surbbd} obeys the dissipation law as:
\begin{align}\label{eq:dissip-law}
\frac{{\rm d}}{{\rm d}t}W(t)=\dot{F}_k + \dot{F}_{_\Phi} + \dot{F}_{w} + \dot{F}_{_\Gamma} + \dot{F}_{_S} + \dot{F}_{_\Lambda},
\end{align}
where the six terms are the viscous dissipation in the bulk fluid ($\dot{F}_{k}$); the diffusion of the bulk surfactants ($\dot{F}_{_\Phi}$); the dissipation at the solid wall ($\dot{F}_w$); the diffusion of the interfacial surfactants ($\dot{F}_{_\Gamma}$); the dissipation due to surfactant adsorption and desorption ($\dot{F}_{_S}$); and the dissipation at the contact points ($\dot{F}_{_\Lambda}$), respectively. They are given by
\begin{subequations}
\begin{align}
\dot{F}_{k}&=-\sum_{i=1,2}\frac{1}{Re}\int_{\Omega_i(t)}2\eta_i\,\norm{D(\vec u)}_F^2\,\dH^2,\qquad\qquad \dot{F}_{_\Phi}= - \frac{1}{We\cdot Da\cdot Pe_{_\Phi}}\int_{\Omega_1(t)}f^{\prime\prime}(\Phi)|\nabla\Phi|^2\,\dH^2,\nn\\
\dot{F}_{w}&=- \sum_{i=1,2}\frac{1}{Re\cdot l_s}\int_{\Sigma_i(t)}\beta_i|u_s|^2\,\ds,\qquad\qquad\qquad \dot{F}_{_\Gamma}= - \frac{1}{We\cdot Pe_{_\Gamma}} \int_{\Sigma(t)}g^{\prime\prime}(\Gamma)\,|\nabla_s\Gamma|^2\,\ds,\nn\\
\dot{F}_{_S}&=- \frac{1}{We}\int_{\Sigma(t)}\Bigl(f'(\Phi) - g'(\Gamma)\Bigr)\,S(\Phi,~\Gamma)\,\ds,\qquad \dot{F}_{_\Lambda}=- \frac{\beta^*}{Re}\left(\dot{x}_l^2 + \dot{x}_r^2\right),\nn
\end{align}
\end{subequations}
where $f'(\Phi) - g'(\Gamma)=E\ln\left(\frac{\lambda\Phi(1-\Gamma)}{\Gamma}\right)$, $f^{\prime\prime}(\Phi)=\frac{E}{\Phi}$ and $g^{\prime\prime}(\Gamma)=\frac{E}{\Gamma(1-\Gamma)}$.
Furthermore, the total mass of the surfactants (rescaled by $\Phi_{\infty}L^2$) is conserved:
\begin{align}
\frac{\rd }{\rd t}\left(\int_{\Omega_1(t)} \Phi\,\dH^2 + Da\int_{\Sigma(t)}\Gamma\;\ds\right) \equiv 0.
\end{align}
 \end{thm}

For the incompressible Navier-Stokes equations \eqref{eq:model12} with boundary conditions \eqref{eq:bd1}-\eqref{eqn:bd6}, we propose an Eulerian weak formulation in \cref{se:weakNV}. For the surfactant transport equations \eqref{eqn:phig} with boundary conditions \eqref{eqn:surbbd}, we propose an ALE weak formulation in conservative form in  \cref{se:weaksurf}.
\subsection{An Eulerian weak formulation for fluid dynamics}
\label{se:weakNV}
We define the following function spaces for the pressure and the fluid velocity, respectively as
\begin{subequations}
\begin{align}
 \label{eq:P}
&\mathbb{P}:=\left\{\psi\in L^2(\Omega):\;
\int_\Omega \psi\,\dH^2=0\right\},\\
\label{eqn:USpace}
&\mathbb{U}:=\left\{\boldsymbol{\omega}\in \left(H^1(\Omega)\right)^2:\;
\boldsymbol{\omega}\cdot\vec n_w =0\;{\rm on}\;\Sigma_1\cup\Sigma_2;\;
\boldsymbol{\omega}=\vec 0\;{\rm on}\;\Sigma_4;\;{\rm and}\;
\boldsymbol{\omega}(-L_x,~y) = \boldsymbol{\omega}(L_x,~y)\right\},
\end{align}
\end{subequations}
with the $L^2$-inner product on $\Omega=\Omega_1(t)\cup\Omega_2(t)$ defined by
\begin{equation}
(\vec u_1,\vec u_2):=\sum_{i=1,2}\left(\vec u_1,~\vec u_2\right)_{\Omega_i(t)}=\sum_{i=1,2}\int_{\Omega_i(t)}\vec u_1\cdot\vec u_2\;\dH^2,\qquad\forall\vec u_1,~\vec u_2\in L^2(\Omega).\nn
\end{equation}

We parameterize the fluid interface as
$\vec X(\alpha, t)=(X(\alpha, t), Y(\alpha, t))^T$,
where $\alpha\in {\mathbb{I}}=[0,1]$, and $\alpha=0, 1$ correspond to the left
and right contact points, respectively.
Define the function spaces for the interface curvature and the interface position as
\begin{subequations}
\label{eqn:kxspace}
\begin{align}
\label{eq:kspace}
& \mathbb{K} := L^2({\mathbb{I}})=\left\{\psi:\;
    \mathbb{I}\rightarrow \mathbb{R}, \;\text{and}
\int_\mathbb{I} |\psi(\alpha)|^2 |\partial_\alpha\vec X| {\rm d}\alpha <+\infty \right\},\\
& \mathbb{X}: = \Big\{ \bmath{g}=(g_1, g_2)^T\in (H^1(\mathbb{I}))^2:\;
      g_2|_{\alpha=0, 1}=0\Big\},
\label{eq:xspace}
\end{align}
\end{subequations}
equipped with the $L^2$-inner product on $\mathbb{I}$
\be
\big(u, v\big)_{\Sigma}:=
\int_{\mathbb{I}}u(\alpha)v(\alpha)\left|\partial_\alpha\vec{X}\right|
{\rm d}\alpha,\qquad \forall\ u, v\in L^2(\mathbb{I}).\nn
\ee

Taking inner product of \eqref{eqn:model1} with $\boldsymbol{\omega}\in \mathbb{U}$,  applying integration by parts and using the boundary \slash interface conditions, we have for the viscous term that
\begin{align}
&\Bigl(\nabla\cdot\mathbf{T}, \boldsymbol{\omega}\Bigr)
=-\Bigl(p, ~\nabla\cdot\boldsymbol{\omega}\Bigr) +
 \frac{2}{Re}\Bigl(\eta D(\vec u), ~D(\boldsymbol{\omega})\Bigr)
 - \Bigl([\mathbf{T}]_1^2\cdot\vec n, ~\boldsymbol{\omega}\Bigr)_\Sigma
+\Bigl(\mathbf{T}\cdot\vec n_w,~\boldsymbol{\omega}\Bigr)_{\Sigma_1\cup\Sigma_2} \nonumber\\
&=-\Bigl( p, ~\nabla\cdot\boldsymbol{\omega}\Bigr)
 + \frac{2}{Re}\Bigl(\eta D(\vec u), ~D(\boldsymbol{\omega})\Bigr)
 +\frac{1}{We}\Bigl(\gamma(\Gamma)\kappa\,\vec n -\nabla_s\gamma(\Gamma), ~\boldsymbol{\omega}\Bigr)_\Sigma + \frac{1}{Re\cdot\,l_s}\Bigl(\beta\,u_s, ~\omega_s\Bigr)_{\Sigma_1\cup\Sigma_2},
\label{eqn:immediate1}
\end{align}
where $\eta=\sum_{i=1,2}\eta_i\,\chi_{_{\Omega_i(t)}}$, $\beta = \sum_{i=1,2}\beta_i\,\chi_{_{\Sigma_i(t)}}$ with $\chi$ being the characteristic function, and $\omega_s = \boldsymbol{\omega}\cdot\vec t_w$.

For the curvature term in Eq.~\eqref{eq:cv}, we can rewrite it as
$\kappa\,\vec n = -\partial_{ss}\vec X$. Multiplying this equation
by a test function $\bmath{g}=(g_1,~g_2)^T\in \mathbb{X}$, and integrating it over $\Sigma(t)$ followed by integration by parts yields
\begin{align}
0 &=\Bigl(\kappa \vec n, ~\bmath{g}\Bigr)_\Sigma
 -\Bigl(\partial_s\vec X, ~\partial_s\bmath{g}\Bigr)_\Sigma
+(\bmath{g}\cdot\vec t)\Big|_{\alpha=0}^{\alpha=1}\nonumber\\
%
%
&=\Bigl(\kappa \vec n, ~\bmath{g}\Bigr)_\Sigma
-\Bigl(\partial_s\vec X, ~\partial_s\bmath{g}\Bigr)_\Sigma
- \beta^*\,Ca\left(\frac{\dot{x}_l\,g_1(0)}{\gamma(\Gamma_l)} + \frac{\dot{x}_r\,g_1(1)}{\gamma(\Gamma_r)} \right)
 + \cos\theta_Y \left(\frac{g_1(1)}{\gamma(\Gamma_r)}-\frac{g_1(0)}{\gamma(\Gamma_r)}\right),
\label{eqn:curvaturefor}
\end{align}
where we have used the fact that $g_2(0)=g_2(1)=0$, $(\vec t\cdot\vec t_w)|_{\alpha=0,1}=\cos\theta_d^{l,r}$ as well as the contact angle condition in \eqref{eqn:bd3}.

From these results, we obtain the weak formulation for the dynamic system (\eqref{eq:model12} with boundary conditions \eqref{eq:bd1}-\eqref{eqn:bd6}) as follows: Given the initial fluid velocity $\vec u_0$ and the interface position $\vec X_0(\alpha)$, find $\vec u(\cdot,~t)\in \mathbb{U}$, $p(\cdot,~t)\in \mathbb{P}$,
$\Sigma(t):=\vec X(\cdot,~t)\in\mathbb{X}$,
and $\kappa(\cdot,~t)\in \mathbb{K}$ such that
\begin{subequations}
\label{eqn:weak1234}
\begin{align}
\label{eqn:weak1}
&\Bigl(\rho\,\frac{\partial\vec u}{\partial t},~\boldsymbol{\omega}\Bigr)
+ \Bigl(\rho\,(\vec u\cdot\nabla)\vec u,~\boldsymbol{\omega}\Bigr)
+\frac{2}{Re}\,\Bigl(\eta D(\vec u),~D(\boldsymbol{\omega})\Bigr) -\Bigl(p,~\nabla\cdot\boldsymbol{\omega}\Bigr)\nonumber\\
& \qquad +\,\frac{1}{We}
\Bigl(\gamma(\Gamma)\kappa\,\vec n - \nabla_s\gamma(\Gamma),~\boldsymbol{\omega}\Bigr)_{\Sigma(t)} +\,\frac{1}{Re\cdot l_s}
\Bigl(\beta\, u_s,~\omega_s\Bigr)_{\Sigma_1(t)\cup\Sigma_2(t)}=\vec 0,
\quad\forall\boldsymbol{\omega}\in \mathbb{U},\\[0.4em]
\label{eqn:weak2}
&\hspace{4cm}\Bigl(\nabla\cdot\vec u,~q\Bigr)=0,
\qquad\forall q\in \mathbb{P}, \\[0.4em]
\label{eqn:weak3}
&\hspace{2cm}\Bigl(\frac{\partial\vec X}{\partial t}\cdot\vec n,~\psi\Bigr)_{\Sigma(t)}
- \Bigl(\vec u\cdot\vec n,~\psi\Bigr)_{\Sigma(t)}=0,\qquad\forall
 \psi\in \mathbb{K}, \\[0.4em]
\label{eqn:weak4}
&\Bigl(\kappa\,\vec n,~\bmath{g}\Bigr)_{\Sigma(t)}-\Bigl
(\partial_s\vec X,~\partial_s\bmath{g}\Bigr)_{\Sigma(t)}-
\beta^*Ca\Bigl(\frac{g_1(1)}{\gamma(\Gamma_r)}\,\dot{x}_r + \frac{ g_1(0)}{\gamma(\Gamma_l)}\,\dot{x}_l\Bigr)\nn\\
&\hspace{2cm} +\cos\theta_Y \left(\frac{g_1(1)}{\gamma(\Gamma_r)} - \frac{g_1(0)}{\gamma(\Gamma_l)}\right)=0,\qquad
\forall\bmath{g}=(g_1,~g_2)^T \in \mathbb{X},
\end{align}
\end{subequations}
where $\rho=\sum_{i=1,2}\rho_i\,\chi_{_{\Omega_i(t)}}$. The interfacial surfactant concentration $\Gamma$ is solved via the weak formulation of \eqref{eqn:phig}, which will be presented in \cref{se:weaksurf}. The above system \eqref{eqn:weak1234} is an extension of the weak formulation in \cite{Zhao19,Zhao20ewod} to the MCL system with surfactants.

\subsection {An ALE weak formulation for surfactant concentrations}
\label{se:weaksurf}

Let $\Omega_1(t)$ be a bounded Lipschitz domain and $\mathcal{O}$ be a reference domain. Take $\left\{\mathcal{A}_t\right\}_{t\in[0,T]}$ as a family of the ALE mappings, i.e., $\mathcal{A}_t(\bmath{\xi})=\vec x(\bmath{\xi},~t)$, where $\bmath{\xi}\in\mathcal{O},\vec x\in\Omega_1(t)$, and assume they satisfy $\mathcal{A}_t\in \left[W^{1,\infty}(\mathcal{O})\right]^2,\, \mathcal{A}_t^{-1}\in\left[W^{1,\infty}(\Omega_1(t))\right]^2,\, \forall t\in[0,~T]$. The domain mesh velocity is defined by
\begin{align}
\vec w(\vec x,~t):= \frac{\partial\vec x(\bmath{\xi},~t)}{\partial t}\Big|_{\bmath{\xi}=\mathcal{A}_t^{-1}(\vec x)}
\end{align}
In this work, we require that on the boundary the mesh velocity satisfies
\begin{align}
\label{eq:ALEBound}
\Bigl(\vec w(\vec x, t) - \vec u(\vec x,t)\Bigr)\cdot\vec n = 0,\quad{\rm on}\;\Sigma(t);\qquad  \Bigl(\vec w(\vec x, t) - \vec u(\vec x,t)\Bigr)\cdot\vec n_w = 0,\quad{\rm on}\;\Sigma_1(t),
\end{align}
where $\vec u$ is the fluid velocity. The construction of the ALE mappings will be presented in \cref{se:FEM}.

We define a function space compatible with the ALE mapping by
\begin{align}
\label{eq:ALESpace}
\mathcal{U}:=\Bigl\{\zeta: \bigcup_{t\in[0,T]}\Omega_1(t)\times\{t\}\to\mathbb{R},\quad \zeta = \hat{\zeta}\circ\mathcal{A}_t^{-1},\quad \hat{\zeta}\in H^1(\mathcal{O})\Bigr\}.
\end{align}
Given $\zeta\in\mathcal{U}$, it follows directly from the Reynolds transport formula \eqref{eqn:dderivative} that
\begin{align}
\frac{\rd }{\rd t}\int_{\Omega_1(t)}\zeta\,\Phi\,\dH^2 =\int_{\Omega_1(t)}\frac{\rD(\zeta\,\Phi)}{\rD t} + \zeta\,\Phi\,\nabla\cdot\vec u\,\dH^2=\int_{\Omega_1(t)}\zeta\frac{\rD\Phi}{\rD t} + \Phi\frac{\rD\zeta}{\rD t}\,\dH^2,
\label{eq:wf1}
\end{align}
where we have used the divergence free condition \eqref{eqn:model2}. Using the chain rule, the derivative with respect to the reference frame can be written as
\begin{align}
\left.\frac{\partial\zeta}{\partial t}\right|_{\mathcal{O}} := \frac{\partial}{\partial t}\left(\zeta(\cdot, t)\circ\mathcal{A}_t\right) = \frac{\partial\zeta}{\partial t} + \vec w\cdot\nabla\zeta.
\label{eq:alef}
\end{align}
On the other hand, since the reference frame is time independent, i.e., $\hat{\zeta}=\zeta\circ\mathcal{A}_t$ is independent of $t$ for $\forall \zeta\in\mathcal{U}$, we have
\begin{equation}
  \left.\frac{\partial\zeta}{\partial t}\right|_{\mathcal{O}}=0.
  \label{eq:alef2}
\end{equation}
Combining \eqref{eq:alef}, \eqref{eq:alef2} and \eqref{eqn:mderivative}, we obtain
\begin{align}
\frac{\rD\zeta}{\rD t} =(\vec u-\vec w)\cdot\nabla\zeta.
\label{eq:Dzeta}
\end{align}


Taking the inner product of Eq.~\eqref{eqn:phis} over $\Omega_1(t)$ with $\zeta\in \mathcal{U}$, we obtain
\begin{align}
\Bigl(\frac{\partial\Phi}{\partial t} + \vec u\cdot\nabla\Phi,~\zeta\Bigr)_{\Omega_1(t)}+ \frac{1}{Pe_{_\Phi}}\Bigl(\nabla \Phi,~\nabla \zeta \Bigr)_{\Omega_1(t)} +Da\Bigl(S(\Phi,\Gamma),~\zeta\Bigr)_{\Sigma(t)}=0,
\label{eq:bulkin}
\end{align}
where we have applied integration by parts as well as the boundary conditions in \eqref{eqn:surbbd}.
Combining \eqref{eq:wf1}, \eqref{eq:Dzeta} and \eqref{eq:bulkin}, we arrive at
\begin{align}
\frac{\rd }{\rd t}\Bigl(\Phi,~\zeta\Bigr)_{\Omega_1(t)}-\Bigl(\Phi\,(\vec u-\vec w),~\nabla\zeta\Bigr)_{\Omega_1(t)} + \frac{1}{Pe_{_\Phi}}\Bigl(\nabla \Phi,~\nabla \zeta \Bigr)_{\Omega_1(t)} +Da\Bigl(S(\Phi,\Gamma),~\zeta\Bigr)_{\Sigma(t)}=0,
\label{eq:aleweakbulk}
\end{align}
where integration by parts and the boundary condition \eqref{eq:ALEBound} are used.

On the fluid interface $\Sigma(t)$, we denote $\mathcal{O}_{_\Sigma}:=\mathcal{A}_t^{-1}(\Sigma(t))\subset\mathcal{O}$
as the reference domain of $\Sigma(t)$. Similarly, we define the function space compatible with the ALE mapping on $\Sigma(t)$ as
\begin{align}
\mathcal{U}_{_\Sigma}:=\Bigl\{\phi: \bigcup_{t\in[0,T]}\Sigma(t)\times\{t\}\to\mathbb{R},\quad \phi= \hat{\phi}\circ\mathcal{A}_t^{-1},\quad \hat{\phi}\in H^1(\mathcal{O}_{_\Sigma})\Bigr\}.
\end{align}
Given $\phi\in \mathcal{U}_{_\Sigma}$, an application of the Reynolds transport formula on the surface in \eqref{eqn:sderivative} gives rise to (see also \cite{Elliott2012})
\begin{align}
\label{eqn:surfacederivative}
\frac{{\rm d}}{{\rm d}t}\Big(\Gamma,~\phi\Big)_{\Sigma(t)} = \Bigl(\frac{\rD\Gamma}{\rD t},~\phi\Bigr)_{\Sigma(t)} + \Bigl(\Gamma,~\frac{\rD\phi}{\rD t}\Bigr)_{\Sigma(t)} + \Bigl(\Gamma\,\nabla_s\cdot\vec u,~\phi\Bigr)_{\Sigma(t)}.
\end{align}
%

A similar result as \eqref{eq:Dzeta} holds on $\Sigma(t)$,
\begin{align}
\frac{\rD\phi}{\rD t} = (\vec u -\vec w_{_\Sigma})\cdot\nabla_s\phi,
\label{eq:Dphi}
\end{align}
where $\vec w_{_\Sigma} =\vec w|_{_{\Sigma(t)}}$.
To see this, let us properly extend $\phi\in \mathcal{U}_{_\Sigma}$ to $\overline{\phi}\in\mathcal{U}$ such that $\left.\overline{\phi}(\vec x, t)\right|_{\Sigma} = \phi$ (this extension can be obtained as the preimage under $\mathcal{A}_t$ of a proper extension of $\hat{\phi}$ in $H^1(\mathcal{O})$). Applying \eqref{eq:Dzeta} to $\zeta=\overline{\phi}$, and restricting the result on $\Sigma(t)$, we obtain
\begin{align}
\frac{\rD\phi}{\rD t} = (\vec u - \vec w_{_\Sigma})\cdot(\mathcal{P}_{_\Sigma}\nabla\overline{\phi})=(\vec u -\vec w_{_\Sigma})\cdot\nabla_s\phi,
\end{align}
where we have used the fact that $(\vec u - \vec w)\cdot\vec n=0$ on $\Sigma$.


Now multiplying \eqref{eqn:gammas} with $\phi\in\mathcal{U}_{_\Sigma}$ followed by an integration over $\Sigma(t)$, and making use of \eqref{eqn:surbd3}, \eqref{eqn:surfacederivative} and \eqref{eq:Dphi}, we arrive at
\begin{equation}
\label{eq:aleweakinter}
\frac{{\rm d}}{{\rm d}t}\Bigl(\Gamma,~\phi\Bigr)_{\Sigma(t)} - \Bigl(\Gamma\,(\vec u - \vec w_{_\Sigma}),~\nabla_s\phi\Bigr)_{\Sigma(t)} + \frac{1}{Pe_{_\Gamma}}\Bigl(\nabla_s\Gamma,~\nabla_s\phi\Bigr)_{\Sigma(t)} = \Bigl(S(\Phi,\Gamma),~\phi\Bigr)_{\Sigma(t)}.
\end{equation}

Collecting results in \eqref{eq:aleweakbulk} and \eqref{eq:aleweakinter}, we obtain the weak formulation  for the surfactant transport equations \eqref{eqn:phig} with boundary conditions \eqref{eqn:surbbd} as follows: Given the surfactant concentrations $\Phi_0=\Phi(\cdot,0)$ and $\Gamma_0:=\Gamma(\cdot,0)$, for all $t\in(0,~T]$ with terminal time $T$, we find $\Phi(\cdot,t)\in H^1(\Omega_1(t))$
and
$\Gamma(\cdot,t)\in H^1(\Sigma(t))$
such that $\forall \zeta\in \mathcal{U}$ and $\phi\in\mathcal{U}_{_\Sigma}$, the following two equations hold
\begin{subequations}
\label{eqn:weakbisurf}
\begin{align}
\label{eq:weakbulksurf}
&\frac{\rd }{\rd t}\Bigl(\Phi,~\zeta\Bigr)_{\Omega_1(t)} + \frac{1}{Pe_{_\Phi}}\Bigl(\nabla \Phi,~\nabla \zeta \Bigr)_{\Omega_1(t)}- \Bigl(\Phi\,(\vec u - \vec w),~\nabla\zeta\Bigr)_{\Omega_1(t)}+\,Da\Bigl(S(\Phi,\Gamma),~\zeta\Bigr)_{\Sigma(t)}=0,\\[0.4em]
&\frac{{\rd}}{\rd t}\Bigl(\Gamma,~\phi\Bigr)_{\Sigma(t)} + \frac{1}{Pe_{_\Gamma}}\Bigl(\nabla_s\Gamma,~\nabla_s\phi\Bigr)_{\Sigma(t)}- \Bigl(\Gamma\,(\vec u- \vec w_{_\Sigma}),~\nabla_s\phi\Bigr)_{\Sigma(t)}- \Bigl(S(\Phi,\Gamma),~\phi\Bigr)_{\Sigma(t)}=0.
\label{eq:weakintersurf}
\end{align}
\end{subequations}
The ALE formulation for both the bulk surfactants and interfacial surfactants are in conservative form. By choosing $\zeta=\chi_{_{\Omega_1(t)}}$ in \eqref{eq:weakbulksurf} and $\phi=\chi_{_{\Sigma(t)}}$
in \eqref{eq:weakintersurf}, and combining the two equations, we have
\begin{align}
\frac{\rd }{\rd t}\Bigl(\Phi,~1\Bigr)_{\Omega_1(t)} + Da\,\frac{\rd }{\rd t}\Bigl(\Gamma,~1\Bigr)_{\Sigma(t)} = 0,
\end{align}
which implies that the total mass of the surfactants is exactly conserved within the weak formulation. We note that in \cite{Barrett2015insoluble,Barrett2015soluble}, a  formulation that is similar to \eqref{eq:weakintersurf} was derived for the interfacial surfactants.

\section{Finite element approximations}
\label{se:FEM}
Based on the weak formulations \eqref{eqn:weak1234} and \eqref{eqn:weakbisurf}, we present the finite element approximations. Besides, we propose an algorithm to construct the discrete ALE mappings via solving the linear elastic equation.

\subsection{Numerical approximation for the dynamics of two-phase flow}
We uniformly partition the time domain: $[0, T]=\bigcup_{m=1}^M [t_{m-1}, t_m]$ with $t_m = m\tau,\tau=T/M$ and the reference domain: $\mathbb{I}=\bigcup_{j=1}^{J_{\Sigma}}\mathbb{I}_j$ with $\mathbb{I}_j=[\alpha_{j-1},~\alpha_j], \alpha_j = jh, h=1/J_{_\Sigma}$. We then define the following finite element spaces to approximate the function spaces $\mathbb{K}$ and $\mathbb{X}$ in \eqref{eqn:kxspace}:
\begin{subequations}
\begin{align}
&\mathbb{K}^h:=\Big\{\psi\in C(\mathbb{I}): \,\psi|_{\mathbb{I}_j}\in
\mathcal{P}_1(\mathbb{I}_j),\quad \forall\, j = 1,2,\cdots, J_{\Sigma}\Big\},\\
& \mathbb{X}^h:=\Big\{ \bmath{g}=(g_1, g_2)^T \in (C(\mathbb{I}))^2: \,
\bmath{g}|_{\mathbb{I}_j}\in (\mathcal{P}_1(\mathbb{I}_j))^2,  \quad
\forall\, j = 1,2,\cdots, J_{\Sigma};\quad g_2|_{\alpha=0, 1} =0\Big\},
\end{align}
\end{subequations}
where $\mathcal{P}_1(\mathbb{I}_j)$ denotes the space of the polynomials of degree at most 1 over $\mathbb{I}_j$.

For $m\geqslant 0$, denote by $\Sigma^m:=\vec X^m(\cdot)=(X^m,~Y^m)^T\in\mathbb{X}^h$ the numerical approximation of $\vec X(\cdot,~t_m)$. Then $\Sigma^m$ is a polygonal curve consisting of connected line segments. We approximate the inner product $(u,v)_{_{\Sigma(t_m)}}$  by either the Simpson's rule  or the trapezoidal rule  as
\begin{subequations}
\begin{align}
\label{eqn:fullsimpson}
&\big(u, v\big)_{\Sigma^m}:=\frac{1}{6}\sum_{j=1}^{J_\Sigma}\left|\vec h_j^m\right|\Big[\big(u\cdot v\big)(\alpha_{j-1}^+)
+4\big(u\cdot v\big)(\alpha_{j-\frac{1}{2}})+\big(u\cdot v\big)(\alpha_j^-)\Big],\\
\label{eqn:fullmassnorm}
&\big(u, v\big)_{\Sigma^m}^h:=\frac{1}{2}\sum_{j=1}^{J_\Sigma}\left|\vec h_j^m\right|\Big[\big(u\cdot v\big)(\alpha_{j-1}^+)
+\big(u\cdot v\big)(\alpha_j^-)\Big],
\end{align}
\end{subequations}
where $\vec h_j^m=\vec{X}^m(\alpha_{j})-
\vec{X}^m(\alpha_{j-1})$, and $u(\alpha_j^\pm)=\lim\limits_{\alpha\to \alpha_j^\pm} u(\alpha)$ are the one-sided limits. The unit tangential and normal vectors are step functions over $\mathbb{I}$ and can be calculated as
\begin{align}
\vec t_j^m=\vec t^m|_{\mathbb{I}_j}=\frac{\vec h_j^m}{|\vec h_j^m|},\quad \vec n^m = \left(\vec t^m\right)^\perp,\qquad 1\leqslant j \leqslant J_{_\Sigma},\quad 0\leqslant m\leqslant M,
\end{align}
where $(\cdot)^\perp$ represents a counter-clockwise rotation of a vector in $\mathbb{R}^2$.

Let $\mathscr{T}^m=\bigcup_{j=1}^N\,\overline{o}_j^m$ be a regular partition of $\Omega$ with $N$ mutually disjoint non-degenerate triangles and a collection of $K$ vertices $\left\{\bmath{a}_k^m\right\}_{k=1}^{K}$. We use a fitted mesh such that the interface mesh $\Sigma^m$ is fitted to the bulk mesh $\mathscr{T}^m$. In other words, the line segments of $\Sigma^m$ are edges of triangles from $\mathscr{T}^m$. We then define the finite element spaces
\begin{subequations}
\begin{align}
\label{eqn:fullskspace}
S_k^m&:=\left\{\psi\in C(\bar{\Omega}):\psi|_{o_j^m}\in \mathcal{P}_k(o_j^m),\;\forall j=1,\cdots,N\right\},\\
\label{eqn:fulls0space}
S_0^m&:=\left\{\psi\in L^2(\Omega):\psi|_{o_j^m}\in \mathcal{P}_0(o_j^m),\;\forall j=1,\cdots,N\right\},
\end{align}
\end{subequations}
where for positive integer $k$, $\mathcal{P}_k(o_j^m)$ denotes the spaces of polynomials with degree at most $k$ on $o_j^m$, and $\mathcal{P}_0(o_j^m)$ denotes the spaces of piecewise constant functions on $o_j^m$. Denote $\mathbb{U}^m$ and $\mathbb{P}^m$ as the finite element spaces for the numerical solution of the velocity and pressure at $t=t_m$. In this work, we choose them as
\begin{equation}
\label{eqn:fullspaceUP}
\left(\mathbb{U}^m,~\mathbb{P}^m\right)=\left((S_2^m)^2\cap\mathbb{U},~(S_1^m+S_0^m)\cap\mathbb{P}\right),
\end{equation}
which satisfies the inf-sup stability condition \cite{Zhao19,Boffi12}. We note that $\Sigma^m$ divides the domain $\Omega$ into two disjoint subdomains $\Omega_1^m$ and $\Omega_2^m$ with corresponding triangular meshes $\mathscr{T}_1^m$ and $\mathscr{T}_2^m$. This gives the approximation of the density, viscosity and friction coefficient: $\rho^m=\sum_{i=1,2}\rho_i\chi_{_{\Omega_i^m}},\; \eta^m=\sum_{i=1,2}\eta_i\chi_{_{\Omega_i^m}},\;\beta^m=\sum_{i=1,2}\beta_i\chi_{_{\Sigma_i^m}}$.

Now we propose the full discreteization of the weak formulation \eqref{eqn:weak1234} as follows: Let $\Sigma^0:=\vec X^0(\cdot )\in \mathbb{X}^h$ and $\mathscr{T}^0$ be the initial discretization of $\Omega$, and $\vec u^0=\pi_2^0\vec u_0\in \mathbb{U}^0$ be the initial fluid velocity.
For $m \geqslant 0$, we seek $\vec u^{m+1}\in \mathbb{U}^m$, $p^{m+1}\in\mathbb{P}^m$,
$\vec X^{m+1}\in \mathbb{X}^h$, and $\kappa^{m+1}\in\mathbb{K}^h$ by solving
the following equations:
\begin{subequations}
\label{eqn:full1234}
\begin{align}
\label{eqn:full1}
&\Bigl(\rho^{m}\frac{\vec u^{m+1}-\pi_2^m\vec u^m}
{\tau},~\boldsymbol{\omega}^h\Bigr) +\Bigl(\rho^m(\pi_2^m\vec u^m\cdot\nabla)
\vec u^{m+1},~\boldsymbol{\omega}^h\Bigr)
+\frac{2}{Re} \Bigl(\eta^m D(\vec u^{m+1}),~D(\boldsymbol{\omega}^h)\Bigr)\nonumber\\
& \qquad - \Bigl(p^{m+1},~\nabla\cdot\boldsymbol{\omega}^h\Bigr)+\frac{1}{We}\Bigl(\gamma^m\kappa^{m+1}\vec n^m-\nabla_s\gamma^m,~\boldsymbol{\omega}^h
\Bigr)_{\Sigma^m}\nonumber\\
&\hspace{3cm}+ \frac{1}{Re\cdot l_s}\Bigr(\beta^m\,u_s^{m+1},
~\omega_s^h\Bigr)_{\Sigma_1^m\cup\Sigma_2^m}=\vec 0,\quad\forall\boldsymbol{\omega}^h
\in \mathbb{U}^m,\\[0.5em]
\label{eqn:full2}
&\hspace{3cm}\Bigl(\nabla\cdot\vec u^{m+1},~q^h\Bigr)=0,
        \qquad\forall q^h\in \mathbb{P}^m,\\[0.5em]
\label{eqn:full3}
&\hspace{1cm} \Bigl(\frac{\vec X^{m+1}-\vec X^m}{\tau}\cdot\vec n^m,
~\psi^h\Bigr)_{\Sigma^m}^h - \Bigl(\vec u^{m+1}\cdot\vec n^m,~\psi^h\Bigr)
_{\Sigma^m}=0,\quad\forall \psi^h\in\mathbb{K}^h,\\[0.5em]
&\Bigl(\kappa^{m+1}\,\vec n^m,~\bmath{g}^h\Bigr)_{\Sigma^m}^h
-\Bigl(\partial_s\vec X^{m+1},~\partial_s\bmath{g}^h\Bigr)_{\Sigma^m}
 +\cos\theta_Y \left(\frac{g_1^h(1)}{\gamma^m_r} - \frac{g_1^h(0)}{\gamma^m_l}\right)\nonumber\\
&\hspace{2em}
-\,\frac{\beta^* Ca}{\tau}\left(\frac{\left(x^{m+1}_r- x^m_r\right)}{\gamma^m_r}g_1^h(1)
+\frac{\left(x^{m+1}_l-x_l^m\right)}{\gamma_l^m} g_1^h(0)\right)=0,
\quad\forall\bmath{g}^h=(g_1^h,~g_2^h)^T\in\mathbb{X}^h.
\label{eqn:full4}
\end{align}
\end{subequations}
Here in the numerical method we have defined
\begin{align*}
 \omega_s^h = \boldsymbol{\omega}^h\cdot\vec t_w,\quad
x_l^m = X^m(0),\quad x_r^m = X^m(1),\quad \gamma^m=\gamma(\Gamma^m),\quad \left.\gamma_{l,r}^m=\gamma^m\right|_{x_{l,r}},
\end{align*}
where $\Gamma^m\in \mathbb{K}^h$ is the interfacial surfactant concentration at $t=t_m$ and its computation will be presented in section \ref{sec:sfnum}. In the above numerical scheme, we have introduced the standard interpolation operator $\pi^m_2: (C(\Omega))^2\to\mathbb{U}^m$. This is because $\vec u^m$ is obtained on the mesh $\mathscr{T}^{m-1}$, it needs to be interpolated from $\mathscr{T}^{m-1}$ to $\mathscr{T}^m$ in order to be used to solve for $\left(\vec u^{m+1},~p^{m+1},~\vec X^{m+1},~\kappa^{m+1}\right)$ on $\mathscr{T}^m$.
For any $f\in\mathbb{K}^h$, we compute its derivative with respect to the arclength parameter on $\Sigma^m$ as $\partial_s f=\frac{\partial_\alpha f}{|\partial_\alpha\vec X^m|}$. To be sufficiently accurate, we apply Simpson's rule \eqref{eqn:fullsimpson} for the numerical integrations of all inner products on the interface $\Sigma^m$ except for the first terms in \eqref{eqn:full3} and \eqref{eqn:full4}. The first terms in \eqref{eqn:full3} and \eqref{eqn:full4} are approximated using trapezoidal rule \eqref{eqn:fullmassnorm} in order to make the interfacial mesh equi-distributed \cite{Zhao20}. Therefore, in our numerical experiments presented below, no re-meshing for the fluid interface is needed during the simulation.


\subsection{The discrete ALE mappings and computation of surfactant concentrations}\label{sec:sfnum}

For $m\geqslant 0$, the discretization of the $\Omega^m$ is given by $\mathscr{T}^m:=\cup_{j=1}^N\overline{o}_j^m$. Then  $\mathscr{T}^{m+1}$ is generated based on $\mathscr{T}^m$ and the displacement of the mesh points on the interface. Specifically,  we keep the mesh connectivity and topology unchanged and update the vertices of the triangular mesh according to
\begin{equation}
\label{eqn:movingmesh1}
\bmath{a}_k^{m+1} = \bmath{a}_k^m + \boldsymbol{\Psi}|_{\bmath{a}_k^m},\nn
\end{equation}
where $\{\bmath{a}_k^m\}_{k=1}^{K}$ are the vertices of $\mathscr{T}^m$, and $\boldsymbol{\Psi}(\vec x)=(\Psi^1,~\Psi^2)^T\in\left(S_1^m\right)^2$ is the displacement vector. Suppose we have solved for $\vec X^{m+1}$ on $\mathscr{T}^m$ from \eqref{eqn:full1234}. This gives the numerical solution for the interface at $t=t_{m+1}$, and thus the displacement vector $\boldsymbol{\Psi}=\vec X^{m+1}-\vec X^m$ on $\Sigma^m$. The displacement of the vertices on the boundary $\Sigma_1^m\cup\Sigma_2^m$ is $\boldsymbol{\Psi}=(\Psi^1(x),~0)^T$, where $\Psi^1(x)$ is a piecewise linear function as
\begin{equation}   \label{eq:eta1}
\Psi^1(x)= \left\{
\begin{array}{l}
\frac{\Delta x_l^m (x+L_x)}{x_l^m+L_x},\quad -L_x\leqslant x<x_l^m, \vspace{0.15cm}\\
\frac{\Delta x_l^m(x-x_r^m)}{x_l^m-x_r^m} +\frac{\Delta x_r^m(x-x_l^m)}{x_r^m-x_l^m},
\quad x_l^m\leqslant x\leqslant x_r^m,\vspace{0.15cm} \\
\frac{\Delta x_r^m(x-L_x)}{x_r^m-L_x},\quad x_r^m<x\leqslant L_x,
\end{array}\right.
\end{equation}
with $\Delta x_{l,r}^m:=x_{l,r}^{m+1} - x_{l,r}^m$.
We then compute the displacement of the internal vertices by solving the linear elasticity equation
\begin{equation}
\label{eqn:movingmesh}
\nabla\cdot
\Bigl[\lambda(\vec x)\Bigl(\nabla\boldsymbol{\Psi} + (\nabla\boldsymbol{\Psi})^T+(\nabla\cdot\boldsymbol{\Psi})\mathbf{I}\Bigr)\Bigr]=\vec 0,
\end{equation}
on $\mathscr{T}^m$ with $\mathcal{P}_1$ Lagrange element, where the homogeneous Dirichlet boundary conditions are prescribed on $\Sigma_3^m\cup\Sigma_4^m$, and $\lambda(\vec x)$ is defined as
\begin{equation}
\lambda(\vec x)|_{o_i^m}:=1 + \frac{\max_{j=1}^N|o_j^m|-\min_{j=1}^N|o_j^m|}{|o_i^m|},\quad 1\leqslant i\leqslant N,\nn
\end{equation}
and it is used to limit the distortion of small elements.

For $m\geqslant 0$, it is natural to assume $\Omega_1^{m+1}$ as the ALE reference domain. We then construct the discrete ALE mapping as
\begin{align}
\mathcal{A}_{h,t}(\vec x):=\sum_{j\in\mathcal{I}_1}\phi_j(\vec x)\left(\frac{t_{m+1}-t}{\tau}\,\bmath{a}_{j}^m + \frac{t-t_m}{\tau}\,\bmath{a}_{j}^{m+1}\right),\quad t_m\leqslant t\leqslant t_{m+1},\quad \vec x\in\Omega_1^{m+1},
\end{align}
where $\mathcal{I}_1:=\left\{j:\;\bmath{a}_j^{m+1}\in\Omega_1^{m+1}\right\}$, and $\phi_j(\vec x)$ is the nodal basis function of $S_1^{m+1}$ at node $\bmath{a}_j^{m+1}$. As a consequence, the mesh velocity is a piecewise constant function with respect to
$t$ and can be computed as
\begin{align}
\vec w^{m+1}(\vec x)= \frac{\partial\mathcal{A}_{h,t}(\vec x)}{\partial t} = \sum_{j\in\mathcal{I}_1}\phi_j(\vec x)\,\left(\frac{\bmath{a}_j^{m+1} - \bmath{a}_j^m}{\tau}\right),\qquad t_m\leqslant t\leqslant t_{m+1},\quad \vec x\in\Omega_1^{m+1}.
\end{align}
The interface mesh velocity is $\vec w_{_\Sigma}^{m+1}=\vec w^{m+1}|_{\Sigma^{m+1}}$. We define the finite element space
 \begin{align}
 \mathbb{F}^m&:=\left\{\zeta\in C(\overline{\Omega_1^m}):\;\zeta|_{o_j^m}\in\mathcal{P}_1(o_j^m),\quad\forall o_j^m\in\mathscr{T}_1^m\right\},\qquad m\geqslant 0.
 \end{align}
Since the discrete ALE mapping $\mathcal{A}_{h,t}$ is piecewise linear over $\Omega_1^{m+1}$, therefore $\forall\zeta^h\in\mathbb{F}^{m+1}$, we have $\zeta^h\circ\mathcal{A}_{h,t_m}^{-1}\in \mathbb{F}^m$. For a piecewise linear function $\phi^h$ defined over $\Sigma^{m+1}$, it can also be considered as a linear function defined on the reference domain $\mathbb{I}$, i.e., $\phi^h\in\mathbb{K}^h$.

Now we present the full discretization of the weak formulation \eqref{eqn:weakbisurf} as follows:  Given $\Phi^0\in \mathbb{F}^0$, and $\Gamma^0\in \mathbb{K}^h$, for $m\geqslant 0$, we seek $\Phi^{m+1}\in \mathbb{F}^{m+1}$ and $\Gamma^{m+1}\in \mathbb{K}^h$ by solving the following two equations:
\begin{subequations}
\label{eqn:sf12}
\begin{align}
\label{eqn:sf1}
&\frac{1}{\tau}\Bigl(\Phi^{m+1},~\zeta^h\Bigr)_{\Omega_1^{m+1}} + \frac{1}{Pe_{_\Phi}}\Bigl(\nabla \Phi^{m+1},~\nabla\zeta^h\Bigr)_{\Omega_1^{m+1}} - \Bigl(\Phi^{m+1}\,\left(\pi_2^{m+1}\vec u^{m+1}-\vec w^{m+1}\right),~\nabla\zeta^h\Bigr)_{\Omega_1^{m+1}}\nonumber\\
&\hspace{2cm}=\frac{1}{\tau}\Bigl(\Phi^m,~\zeta^h\circ\mathcal{A}_{h,t_m}^{-1}\Bigr)_{\Omega_1^m}-Da\Bigl(S^{m+\frac{1}{2}},~\zeta^h\Bigr)_{\Sigma^{m+1}}^h,\qquad\forall \zeta^h\in \mathbb{F}^{m+1},\\[0.4em]
&\frac{1}{\tau}\Bigl(\Gamma^{m+1},~\phi^h\Bigr)_{\Sigma^{m+1}}^h + \frac{1}{Pe_{_\Gamma}}\Bigl(\nabla_s\Gamma^{m+1},~\nabla_s\phi^h\Bigr)_{\Sigma^{m+1}} -\Bigl(\Gamma^{m+1}\,(\pi_2^{m+1}\vec u^{m+1} - \vec w_{_{\Sigma}}^{m+1}),~\nabla_s\phi^h\Bigr)_{\Sigma^{m+1}}^h \nonumber\\
&\hspace{2cm} = \,\frac{1}{\tau}\Bigl(\Gamma^m,~\phi^h\Bigr)_{\Sigma^{m}}^h + \Bigl(S^{m+\frac{1}{2}},~\phi^h\Bigr)_{\Sigma^{m+1}}^h,\qquad\forall \phi^h\in\mathbb{K}^h,
\label{eqn:sf2}
\end{align}
\end{subequations}
where $S^{m+\frac{1}{2}}= Bi\,\lambda\,\Phi^{m+1} - Bi\left(\lambda\,\Phi^m\circ\mathcal{A}_{h,t_m}+1\right)\Gamma^{m+1}$ is an approximation to the source term $S(\Phi,~\Gamma)$. Eqs.~\eqref{eqn:sf1} and \eqref{eqn:sf2} form a linear coupled system for $\Phi^{m+1}$ and $\Gamma^{m+1}$.

For the above numerical method \eqref{eqn:sf12}, we can prove:
\begin{thm}[Mass conservation of the surfactants]\label{th:con} For $m\geqslant 0$, let $\Bigl(\Phi^{m+1},~\Gamma^{m+1}\Bigr)$ be the numerical solution of \eqref{eqn:sf12}, then it holds
\begin{align}
\label{eq:SFcon}
\Bigl(\Phi^{m+1},~1\Bigr)_{\Omega_1^{m+1}} + Da\,\Bigl(\Gamma^{m+1},~1\Bigr)_{\Sigma^{m+1}}^h = \Bigl(\Phi^{m},~1\Bigr)_{\Omega_1^{m}} + Da\,\Bigl(\Gamma^{m},~1\Bigr)_{\Sigma^{m}}^h.
\end{align}
\end{thm}
 \begin{proof}
 Choosing $\zeta^h=1$ in \eqref{eqn:sf1} and $\phi^h=1$ in \eqref{eqn:sf2}, then multiplying \eqref{eqn:sf2} with $Da$ and summing up the two equations, we obtain directly \eqref{eq:SFcon}.
 \end{proof}

The overall procedure of the proposed numerical scheme can be summarized as follows: Given the initial discretization of $\Sigma(0)$ and $\Omega(0)$ as $\vec X^0\in\mathbb{X}^h$ and $\mathscr{T}^0$, and the initial velocity $\vec u^0=\pi_2^0\vec u_0$, the initial bulk surfactant concentration $\Phi^0$ and interfacial surfactant concentration $\Gamma^0$, let $m=0$. Then
\begin{itemize}
\item [(1)] solve the linear system \eqref{eqn:full1234} to obtain $\left(\vec u^{m+1},~p^{m+1},~\vec X^{m+1},~\kappa^{m+1}\right)$ on $\mathscr{T}^m$;
\item[(2)] based on $\mathscr{T}^m$ and $\vec X^{m+1}$, construct the new mesh $\mathscr{T}^{m+1}$ via solving the elastic equation \eqref{eqn:movingmesh};
\item [(3)] perform interpolations from $\mathscr{T}^m$ to $\mathscr{T}^{m+1}$ to obtain $\pi_2^{m+1}\vec u^{m+1}$;
\item [(4)] obtain the mesh velocity $\vec w^{m+1}$ and $\vec w^{m+1}_{_\Sigma}$, then solve the linear system \eqref{eqn:sf12} to obtain $\Phi^{m+1}$ and $\Gamma^{m+1}$, and go to step (1) with $m=m+1$.
\end{itemize}

\section{Numerical results}
\label{sec:numer}
In this section, we first present the numerical convergence test of the proposed numerical method \eqref{eqn:full1234} and \eqref{eqn:sf12}. We then report numerical results on the spreading of a droplet contaminated with insoluble\slash soluble surfactants, the effects of the adsorption parameters on the surfactant adsorption\slash desorption as well as some applications.

Unless otherwise stated, we will choose the parameters $\rho_2=0.1$, $\eta_2=0.1$, $\beta_2=0.1$, $\beta^*=0.1$, $Re=10$, $Ca=0.1$ and $l_s=0.1$. The computational domain is  $\Omega=[-1,~1]\times[0,~1]$. Initially, the fluid interface of the droplet is given by a semi-circle: $x^2+y^2=0.4^2,\,y\geqslant 0$, and $\vec u^0=\vec 0$.

\subsection{Convergence test}

\begin{table}[!htp]
\centering
\def\temptablewidth{0.80\textwidth}
\vspace{-12pt}
\caption{Spatial errors of the numerical solution and the rate of convergence for the dynamic of the clean interface (``CI''), the interface contaminated with insoluble surfactants (``IS''), and the interface contaminated with soluble surfactants (``SS''). Upper panel: $t=0.5$. Lower panel: $t=1.5$.}\label{tb:order00}
{\rule{\temptablewidth}{1pt}}
\begin{tabular*}{\temptablewidth}{@{\extracolsep{\fill}}c|cc|cc|cc}
\multirow{2}{*}{$J_{_\Sigma}$} & \multicolumn{2}{c|}{``CI''}
&\multicolumn{2}{c|}{``IS''} &\multicolumn{2}{c}{``SS''}\\ \cline{2-7}
&$e_h(t) $ & order &$e_h(t)$
& order &$e_h(t) $ & order  \\ \hline
$16$ & 1.34E-3 & - &1.48E-3 &-& 1.44E-3 &- \\ \hline
$32$ & 3.57E-4 &1.91 &3.99E-4 &1.89& 3.92E-4 &1.88
\\ \hline
$64$ & 9.65E-5 & 1.89 &1.38E-4 &1.53& 1.26E-4 &1.64
 \end{tabular*}
{\rule{\temptablewidth}{1pt}}
{\rule{\temptablewidth}{1pt}}
\begin{tabular*}{\temptablewidth}{@{\extracolsep{\fill}}c|cc|cc|cc}
$16$ & 1.21E-3 & - &1.21E-3 &-& 1.21E-3 &- \\ \hline
$32$ & 3.08E-4 & 1.97 &3.15E-4 &1.94& 3.10E-4 &1.96
\\ \hline
$64$ & 7.65E-5 & 2.01 &7.62E-5 &2.05& 7.63E-5 &2.02
 \end{tabular*}
{\rule{\temptablewidth}{1pt}}
\end{table}

We start by carrying out numerical simulations using different mesh sizes. Let $\vec X^m(\cdot)$ be the numerical solutions for the interface $\vec X(\cdot, ~t_m)$ computed using mesh size $h$. We approximate the numerical solution in any time intervals by the linear interpolation:
\begin{equation}
\vec X_{h}(\alpha,~t) = \frac{t-t_m}{\tau}\vec X^{m+1}(\alpha) + \frac{t_{m+1}-t}{\tau}\vec X^m(\alpha),\qquad t_m\leqslant t\leqslant t_{m+1}.
\end{equation}
Then we measure the spatial error of the numerical solutions by comparing $\vec X_{h}$ and $\vec X_{\frac{h}{2}}$ using the manifold distance discussed in~\cite{Zhao20}, i.e., the area of the symmetric difference region:
\begin{equation}
e_{h}(t) = \left|\left(\Omega_1^{h}(t)\backslash\Omega_1^{\frac{h}{2}}(t)\right)\cup\left(\Omega_1^{\frac{h}{2}}(t)\backslash\Omega_1^{h}(t)\right)\right|,
\end{equation}
where $\Omega_1^{h}(t)$ represent the region enclosed by the open curve $\vec X_{h}(\alpha, ~t)$ and the substrate line ($x$-axis).

We consider the dynamic of a spreading droplet on a hydrophilic substrate with $\theta_Y=\pi/3$ in three different cases: (1) a clean interface (case ``CI''); (2) the interface contaminated by insoluble surfactants (case ``IS'') with initial datum $\Gamma(\cdot,0)=0.5$; (3) a interface contaminated by soluble surfactants (case ``SS'') with initial data $\Gamma(\cdot,0)=0.4$ on the interface and $\Phi(\cdot,0)=0.6$ in the bulk. The time step is chosen uniformly as $\tau=1\times 10^{-4}$, and other parameters are chosen as: $Bi=0$, $Pe_{_\Gamma}=10$ and $E=0.4$ for case ``IS'', and  $Pe_{_\Phi}=10$, $Pe_{_\Gamma}=10$, $Bi=0.1$, $E=0.4$, $Da=0.1$ and $\lambda=0.1$ for case ``SS''. The numerical solutions are obtained on different meshes with $\left(J_{_\Sigma}, N\right) = (16,258),~(32,998),~(64, 3974),~(128,15812)$. Numerical errors $e_{h}$ for the three different cases are shown in Table~\ref{tb:order00}, where we observe that the convergence rate of the numerical errors can reach about 2.

\begin{figure}[!htp]
\centering
\includegraphics[width=0.75\textwidth]{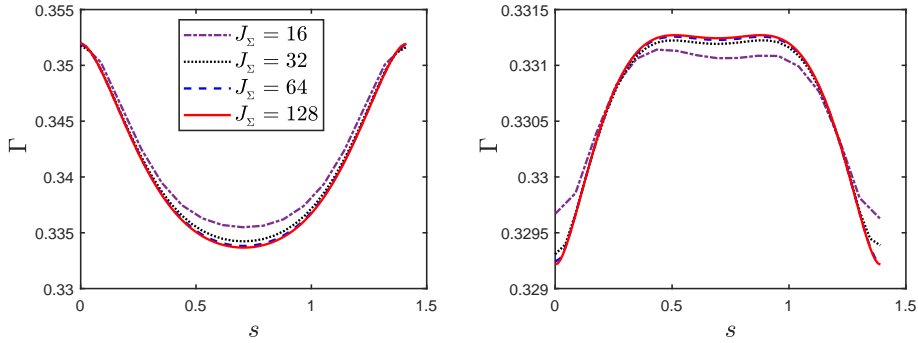}
\caption{The interfacial surfactant concentration versus the arc length for case ``SS'' at time $t=0.5$ (left panel) and $t=1.0$ (right panel) using different mesh sizes. }
\label{Fig:IFs}
\end{figure}

\begin{figure}[!htp]
\centering
\includegraphics[width=0.75\textwidth]{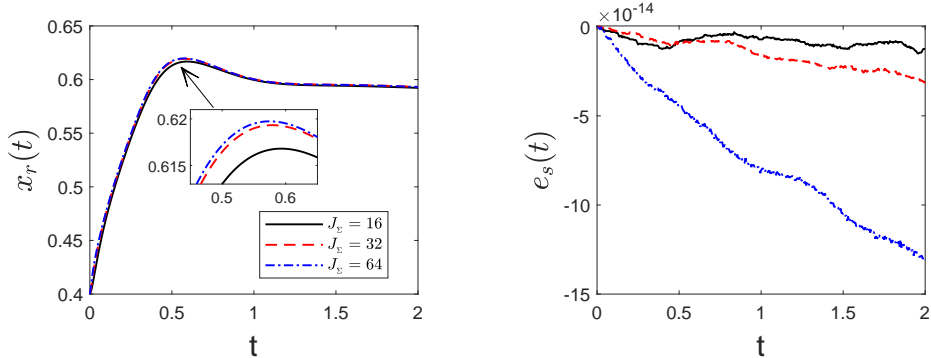}
\caption{The time history of the right contact point position $x_r(t)$ (left panel) and the relative loss of the total mass of the surfactants $e_s(t)$ (right panel).}
\label{Fig:IFs1}
\end{figure}

For case ``SS'', we plot the distribution of the interfacial surfactants versus the arc length parameter under different mesh sizes in Fig.~\ref{Fig:IFs}. We observe the convergence of the numerical solutions as the mesh is refined. We next define the relative loss of the total mass of the surfactants as
\begin{align}
\left.e_s(t)\right|_{t=t_m}=\frac{\int_{\Omega_1^m}\Phi^m\,\dH^2 + Da\int_{\Sigma^m}\Gamma^m\,\ds}{\int_{\Omega_1^0}\Phi^0\,\dH^2 + Da\int_{\Sigma^0}\Gamma^0\,\ds} -1.
\end{align}
The time history of the right contact points position and $e_s(t)$ are depicted in Fig.~\ref{Fig:IFs1}, where we observe the convergence of the contact point position as mesh is refined. Moreover, the total mass of the surfactants is well conserved, which is consistent with Theorem~\ref{th:con}.

\subsection{Applications with insoluble surfactants}
\label{se:Ifnr}

\begin{figure}[!htp]
\centering
\includegraphics[width=0.8\textwidth]{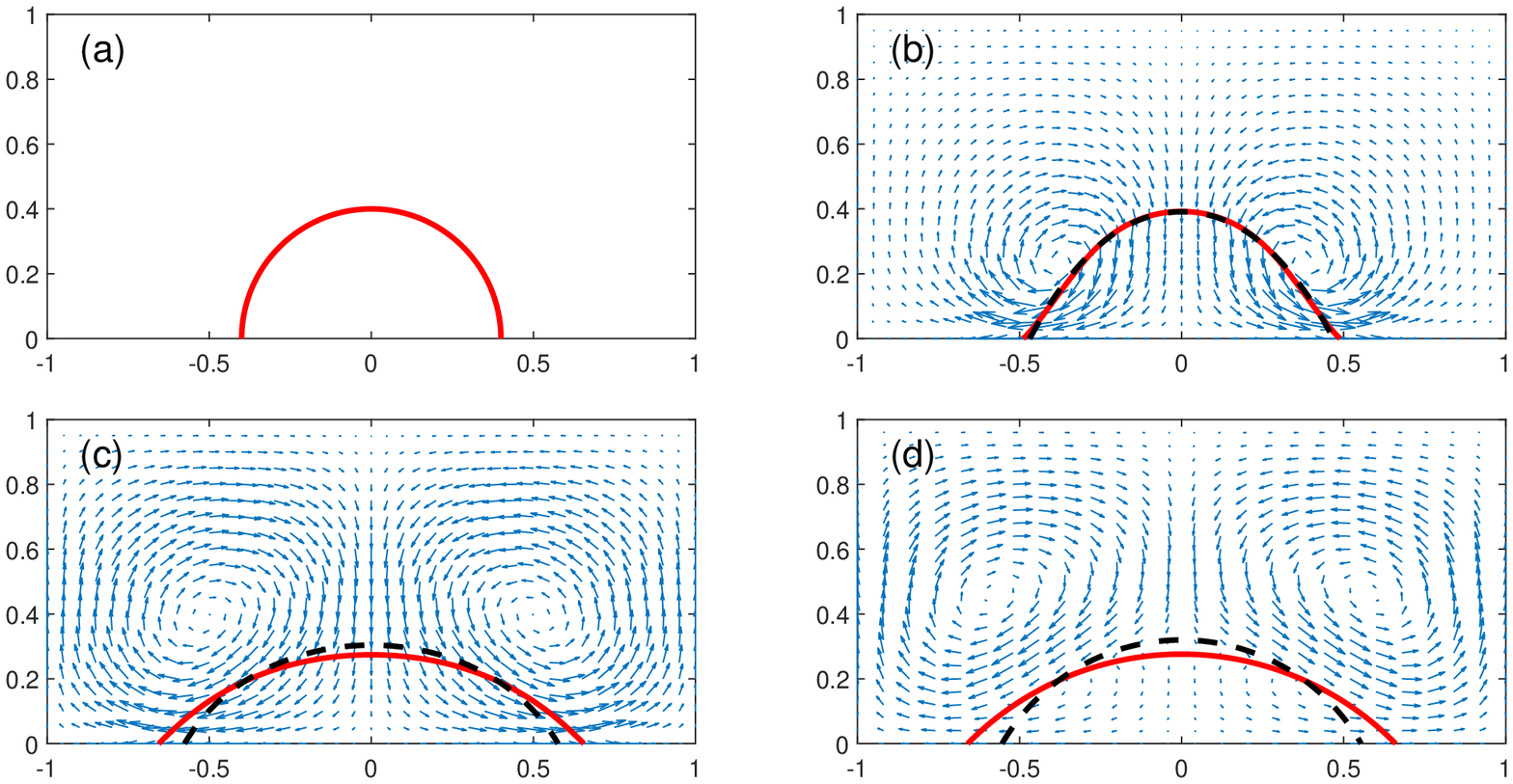}
\caption{Snapshots of the fluid interface (the red solid line) and the velocity field for the droplet on a hydrophilic substrate with $\theta_Y=\pi/3$. (a) $t=0$, $\max_{\vec x\in\Omega}|\vec u|=0$; (b) $t=0.1$, $\max_{\vec x\in\Omega}|\vec u|=0.582$; (c) $t=0.5$, $\max_{\vec x\in\Omega}|\vec u|=0.172$; (d) $t=2.5$, $\max_{\vec x\in\Omega}|\vec u|= 1.86\times10^{-5}$. The black dash line denotes the profile of the clean interface.}
\label{Fig:hydrophi}
\end{figure}

\begin{figure}[!htp]
\centering
\includegraphics[width=0.8\textwidth]{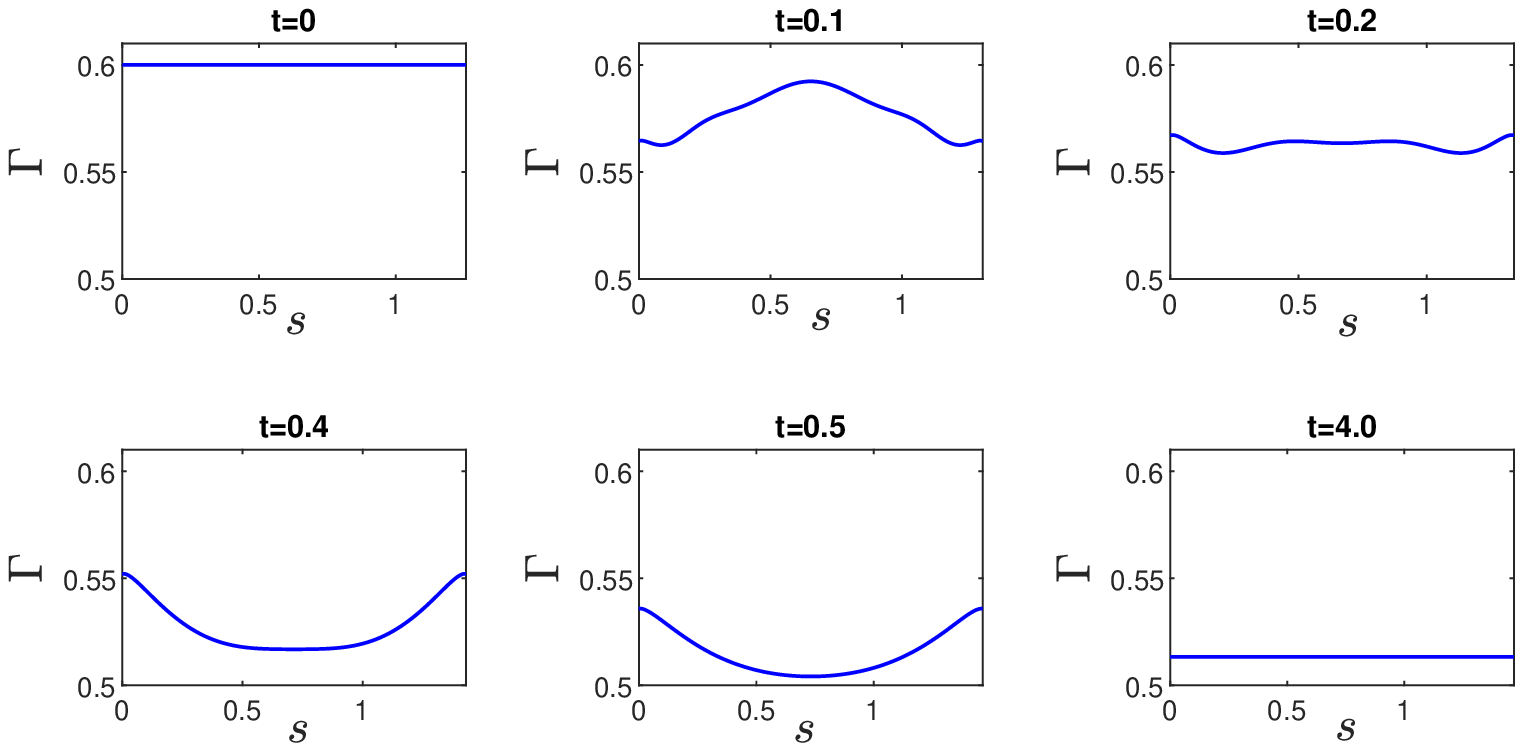}
\caption{Snapshots of the interfacial surfactant concentration  versus the arc length at several times in the hydrophilic case. }
\label{Fig:phisf}
\end{figure}

\begin{figure}[!htp]
\centering
\includegraphics[width=0.70\textwidth]{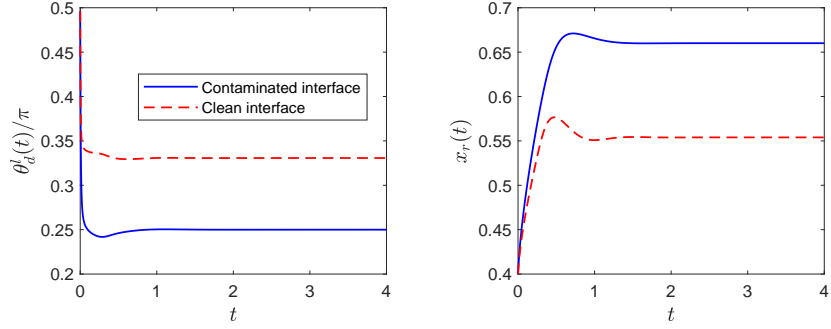}
\caption{The comparison of a contaminated interface with a clean interface on a hydrophilic substrate with $\theta_Y=\pi/3$. Left panel: the evolution of the dynamic contact angle. Right panel: the evolution of the right contact point position. }
\label{Fig:comparsion1}
\end{figure}

\begin{figure}[!htp]
\centering
\includegraphics[width=0.8\textwidth]{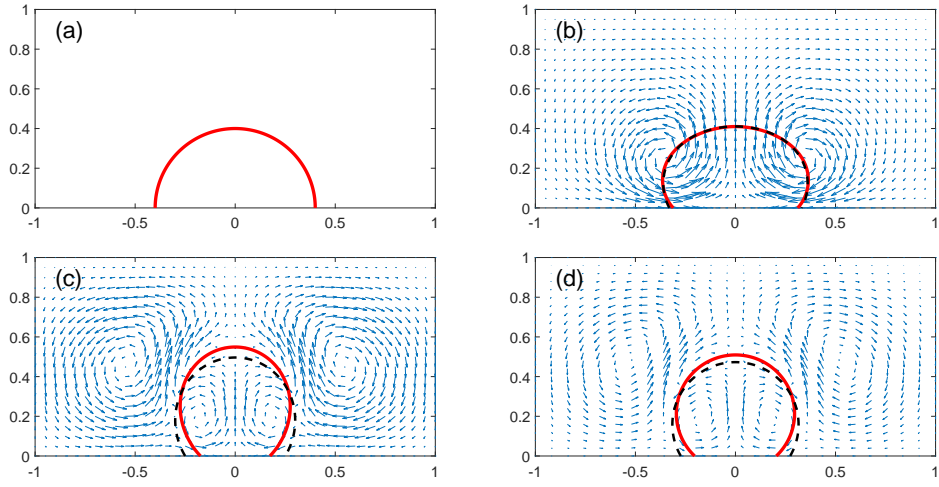}
\caption{Snapshots of the fluid interface (red solid line) and the velocity field on a hydrophobic substrate with $\theta_Y=2\pi/3$. (a) $t=0$, $\max_{\vec x\in\Omega}|\vec u|=0$; (b) $t=0.1$, $\max_{\vec x\in\Omega}|\vec u|=0.581$; (c) $t=0.5$, $\max_{\vec x\in\Omega}|\vec u|=0.088$; (d) $t=2.5$, $\max_{\vec x\in\Omega}|\vec u|=6.54\times 10^{-4}$. The black dash line represents the profile of a clean interface.  }
\label{Fig:hydropho}
\end{figure}

\begin{figure}[!htp]
\centering
\includegraphics[width=0.8\textwidth]{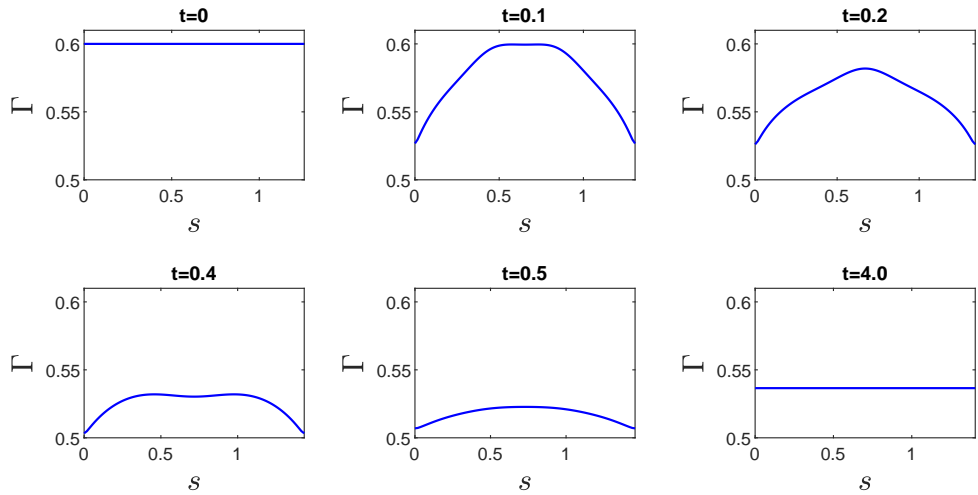}
\caption{Snapshots of the interfacial surfactant concentration  versus the arc length at several moments in the hydrophobic case.}
\label{Fig:hydrophosf}
\end{figure}

We consider the case of insoluble surfactants and examine the dynamics of a spreading droplet on the hydrophilic substrate with $\theta_Y=\pi/3$ and the hydrophobic substrate with $\theta_Y=2\pi/3$. Initially, we set $\Gamma(\cdot,0)=0.6$ uniformly on the interface, and the computational domain is partitioned into $N=12910$ triangles with $K=6600$ vertices. Parameters are chosen as: $Bi=0, \,Pe_{_\Gamma}=10$, $E=0.4$, and $\tau=5\times 10^{-4}$.

Snapshots of the interface and the velocity field at several moments in the hydrophilic case are shown in Fig.~\ref{Fig:hydrophi}, where we plot the corresponding profile of the clean interface as a comparison.  We observe that the contaminated droplet becomes more hydrophilic than the clean one. This is because the interfacial surfactants reduce the surface tension of the contaminated interface, thus decreases the equilibrium contact angle in the hydrophilic case. We also plot the distribution of the interfacial surfactants versus the arc length at several moments in Fig.~\ref{Fig:phisf}. It is found that that at the beginning, the interface evolves fast locally nearby the contact points. The local elongation of the interface around the contact points reduces the local concentration of interfacial surfactants in the neighborhood of the contact points, as shown in Fig.~\ref{Fig:phisf} at $t=0.1$. As the whole interface moves, the surfactants move from the middle part of the interface to the side parts due to the convection and diffusion (see the plots from $t=0.1$ to $t=0.4$). When the contact angle is close to its equilibrium value, the contact point motion slows down, and the interface evolves in a quasi-static manner while keeping a circular arc shape (see also Fig.~\ref{Fig:comparsion1}). The length of the interface increases during the quasi-static movement and thus the surfactant concentration decreases everywhere (the plot in Fig.~\ref{Fig:phisf} at $t=0.5$). Eventually, the surfactants are uniformly distributed along the interface. The comparison of the contact point positions and dynamic contact angles between a contaminated interface and a clean interface is depicted in Fig.~\ref{Fig:comparsion1}.

In the hydrophobic case, we show analogous snapshots for the interface and the velocity field in Fig.~\ref{Fig:hydropho}. We observe the contaminated droplet becomes more hydrophobic than the clean one in this case. The distribution of the interfacial surfactants versus the arc length is shown in Fig.~\ref{Fig:hydrophosf}. Similarly as in the hydrophilic case, the local surfactant concentration near the contact points decreases at the beginning. However, as the whole interface evolves, the surfactants are transported from the side parts of the interface to the middle part (the plots from $t=0.1$ to $t=0.5$). This is because the direction of the velocity field nearby the interface is opposite to that in the hydrophilic case. The resulting convection brings the surfactants from the side parts towards the middle.

\subsection{Langmuir adsorption isotherm}
\label{se:Lanr}

\begin{figure}[!htp]
\centering
\includegraphics[width=0.80\textwidth]{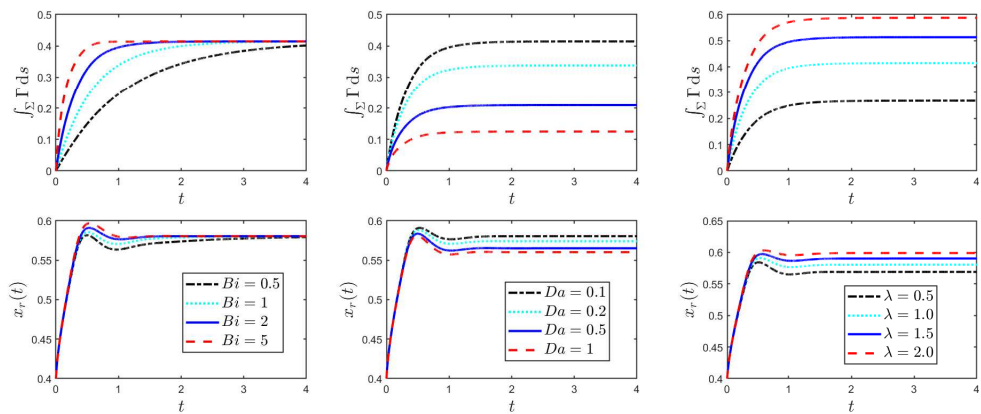}
\caption{The evolution of the total interfacial surfactants (upper panel) and the right contact point (lower panel) under various choices of $Bi, Da$ and $\lambda$. Left panel: $Da=0.1, \lambda=1$; Middle panel: $Bi=2, \lambda=1$; Right panel: $Bi=2, Da=0.1$.}
\label{Fig:binumber}
\end{figure}

We examine the effects of $Bi$, $Da$ and $\lambda$ on the surfactant adsorption\slash desorption to the interface and the contact point dynamics on a hydrophilic substrate with $\theta_Y=\pi/3$. Initially, we choose $\Gamma(\cdot, 0)=0$ on $\Sigma(0)$ and $\Phi(\cdot, 0) = 0.6$ in $\Omega_1(0)$. We fix the bulk and interface P\'ecelt numbers as $Pe_{_\Phi}=10$, $Pe_{_\Gamma}=10$ and surfactant elasticity $E = 0.3$.

The time evolution of the total mass of the interfacial surfactants and the right contact point position using different $Bi$, $Da$ and $\lambda$ are depicted in Fig.~\ref{Fig:binumber}. We observe that the adsorption process is speeded up by increasing $Bi$, $\lambda$ or decreasing $D a$. Besides, we find that the total mass of the interfacial surfactants in equilibrium remains unchanged regardless of the value of $Bi$, and so does the contact point positions. On the contrary, decreasing $Da$ or increasing $\lambda$ increases the total mass of the interfacial surfactants in equilibrium, and thus enhances the droplet spreading. These observations are consistent with our knowledge from the model derivation: the equilibrium constant $\lambda$ tells the relation between the interfacial and the bulk surfactant concentrations in equilibrium, the adsorption depth $Da$ controls the interface adsorption capacity, and $Bi$ gives the speed (or mobility) of the adsorption process.

\begin{figure}[!htp]
\centering
\includegraphics[width=0.75\textwidth]{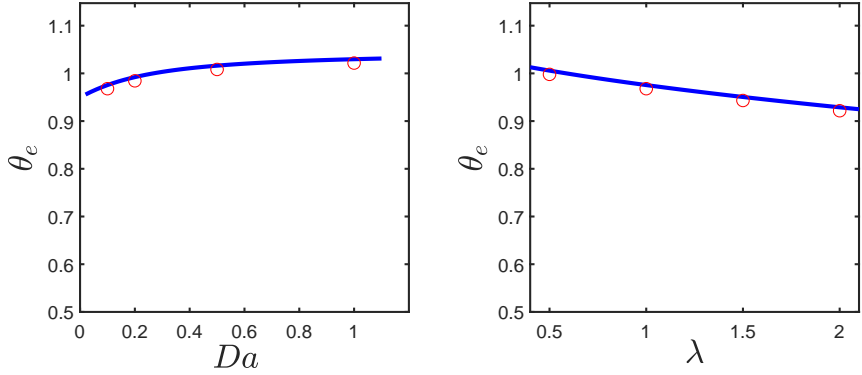}
\caption{Comparison of the numerical equilibrium contact angle $\theta_e$ (discrete red markers) with the analytical solutions (solid blue line) by solving  \eqref{eq:anatheta} using different adsorption depth $Da$ (left panel) and adsorption number $\lambda$ (right panel).}
\label{Fig:ET}
\end{figure}

At equilibrium, both the interfacial and bulk surfactant concentrations are uniform. The fluid interface $\Sigma$ forms a circular arc shape with the radius $R$ and the equilibrium contact angle $\theta_e$. They satisfy the equilibrium conditions
\begin{subequations}
\begin{align}
A_0\Phi + Da\,\Gamma\,|\Sigma| = M_0,\\
\lambda\,\Phi\,(1-\Gamma) = \Gamma,\\
(1+E\ln(1-\Gamma)) \cos\theta_e = \cos\theta_Y,
\label{eq:GT}
\end{align}
\end{subequations}
where $A_0$ and $M_0$ are the area of the droplet and the total mass of the surfactants, respectively. Besides, a straightforward calculation leads to
$R^2(\theta_e-\cos\theta_e\sin\theta_e)=A_0$ and $
2\,R\,\theta_e = |\Sigma|.$
Combining these results, we find that $\theta_e$ satisfies
\begin{align}
\frac{\Gamma}{\lambda\,(1-\Gamma)} A_0 + 2\,Da \Gamma \sqrt{\frac{A_0}{\theta_e-\cos\theta_e\sin\theta_e}}\,\theta_e = M_0.
\label{eq:anatheta}
\end{align}
Therefore, we obtain a quantitative dependence of $\theta_e$ on $Da$ and $\lambda$. The obtained numerical results in Fig.~\ref{Fig:ET} are consistent with the theoretical formulation of $\theta_e$ in \eqref{eq:anatheta}.

\subsection{Applications with soluble surfactants}
\label{se:Apnr}
\begin{figure}[!htp]
\centering
\includegraphics[width=0.8\textwidth]{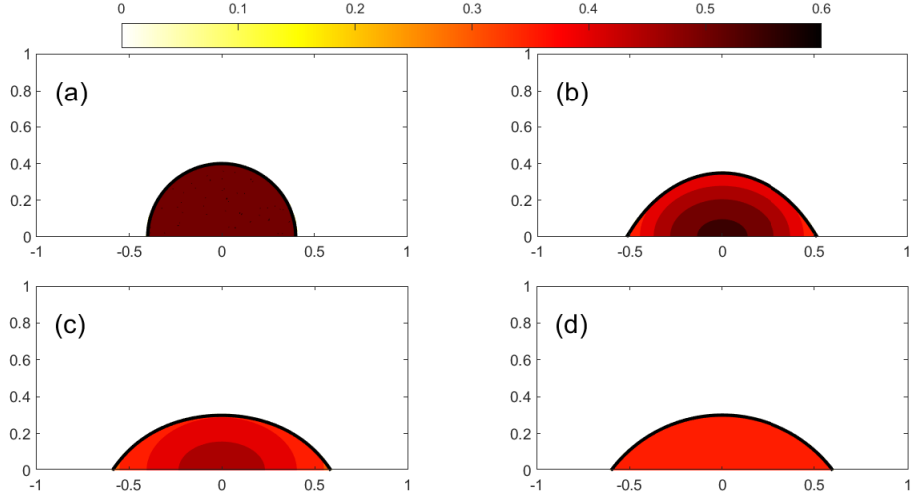}
\caption{Several snapshots of the fluid interface and the contour of the bulk surfactant concentration in the dynamic of a droplet that absorbs surfactants from the bulk domain to the interface. (a) $t=0$; (b) $t=0.2$; (c) $t=0.4$; (d) $t=4.0$. }
\label{Fig:sf1}
\end{figure}

\begin{figure}[!htp]
\centering
\includegraphics[width=0.8\textwidth]{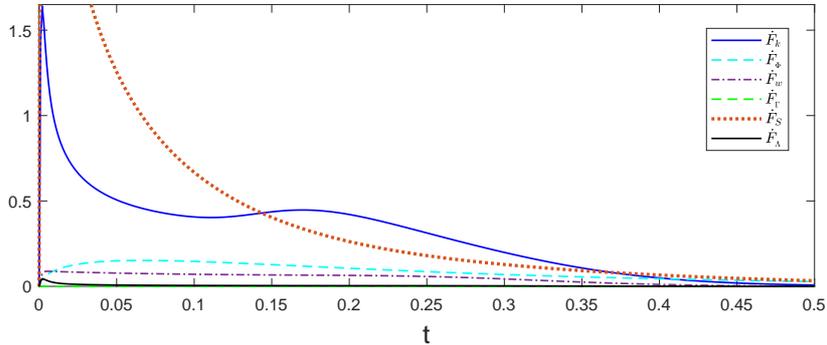}
\caption{Comparison of the different types of energy dissipations (in absolute values). }
\label{Fig:sf3}
\end{figure}

\begin{figure}[!htp]
\centering
\includegraphics[width=0.8\textwidth]{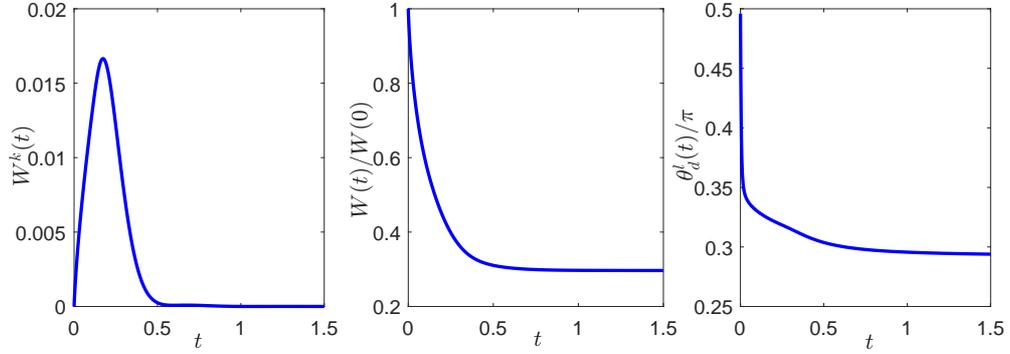}
\caption{The evolution of the fluid kinetic energy $W^k(t)=\frac{1}{2}\int_{\Omega}\rho|\vec u|^2\dH^2$, the normalized total energy $W(t)/W(0)$, and the value of the contact angle. }
\label{Fig:sf2}
\end{figure}

We consider the numerical example in \cref{se:Lanr} with Langmuir adsorption parameters $Bi=2,~Da=0.1$ and $\lambda =2$. The fluid interface together with the contour of the bulk surfactant concentration at several moments are depicted in Fig.~\ref{Fig:sf1}. We observe at the beginning, the bulk surfactants are adsorbed to the fluid interface, thus lowers the nearby concentration, but finally the concentration of the bulk surfactants is uniformly distributed due to the bulk diffusion. During this process, the six types of the dissipation are compared in Fig.~\ref{Fig:sf3}. It is found that at the very beginning the adsorption dissipation $\dot{F}_S$ dominates so that the adsorption process is the most important. Besides, the contact point dynamic and the slip friction is also important. The adsorbed surfactants diffuse on the interface and reduces the surface tension. The gradient in the interfacial surfactant concentration induces a gradient in the surface tension, and thus leads to Marangoni effect. The Marangoni generated flow field in turn transports both the bulk and the interfacial surfactants. The contact point dynamic is also affected due to the change in the surface tension. The interplay among the interface evolution, the bulk and interfacial surfactant diffusions, the adsorption kinetic, and the flow field become dominant in a long period until the system is near the equilibrium state. In Fig.~\ref{Fig:sf2}, we also plot the time evolution of the fluid kinetic energy, the normalized total free energy and the (left) dynamic contact angle. The total free energy is observed to decay with time. This numerically confirms the dissipation law \eqref{eq:dissip-law}. From the evolution of the kinetic energy, the velocity field is of small magnitude after about $t=0.5$ and the system achieves the quasi-static state.

\begin{figure}[!htp]
\centering
\includegraphics[width=0.8\textwidth]{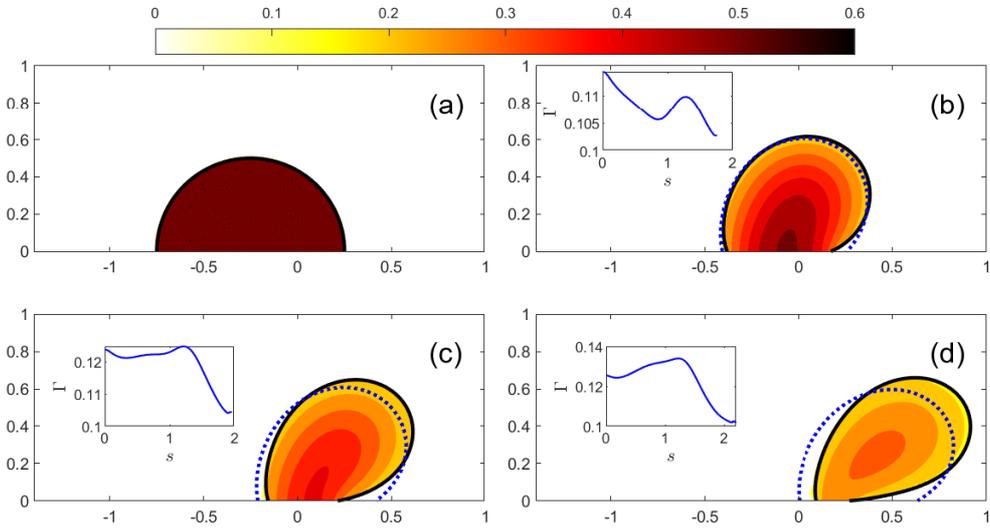}
\caption{Several snapshots of the fluid interface profile (solid black line) together with the contour of the bulk surfactant concentration in the dynamic of a droplet driven by the constant horizontal force on a hydrophobic substrate, where the inset plots show the distribution of the interfacial surfactants. The dotted blue line represents the corresponding clean interface.  (a) $t=0$; (b) $t=0.12$; (c) $t=0.32$; (d) $t=0.60$. }
\label{Fig:Migration}
\end{figure}

We next consider the migration of a droplet on a hydrophobic substrate with $\theta_Y=2\pi/3$. The fluids are driven by the horizontal body force $\vec f=(1,0)^T$. Initially, the fluid interface of the initial droplet is given by a semi-circle $(x+0.25)^2+y^2=0.5^2,\;y\geqslant 0$, and the surfactant concentrations are: $\Gamma(\cdot,0)=0$, $\Phi(\cdot,0)=0.6\chi_{_{\Omega_1(0)}}$. The domain $\Omega$ is partitioned into $N=13237$ triangles with $K=6771$ vertices. Other parameters are chosen as $\rho_1=\rho_2=1$, $\eta_1=\eta_2=1$, $\beta_1=\beta_2=1$, $\beta^*=10$, $Ca=0.02$, $l_s=0.1$, $Pe_{_\Gamma}=Pe_{_\Phi}=10$, $E_s=4.0$, $Da=0.5$, $Bi=2$, $\lambda=1$ and the time step $\tau = 2\times 10^{-4}$.

Several snapshots of the dynamic interface, the distribution of the interfacial surfactants and the contour of the bulk surfactant concentration at different time are depicted in Fig.~\ref{Fig:Migration}. The corresponding profiles of the clean interface (dotted blue line) are also shown as comparison. We observe that both the contaminated droplet and the clean droplet migrate in the direction of the applied force. In both cases, the advancing angle (the right contact angle) is larger than the receding angle (the left contact angle), i.e., $\cos\theta_r<\cos\theta_l<0$. In the case of the contaminated droplet, the interfacial surfactant distribution is no long symmetric and the surfactant concentration nearby the left contact point is larger than that nearby the right one. The accumulation of surfactants near the receding edge lowers the local surface tension more heavily than it does at the advancing edge, in particular, $\gamma(\Gamma_l)<\gamma(\Gamma_r)$. As a consequence, the unbalanced Young's force at the receding contact point (this is positive) is much larger than that at the advancing contact point (this is negative). It can be implied from \eqref{eqn:bd3} that the receding edge moves faster than the advancing edge. This makes the droplet inclined towards advancing edge and further increases the advancing contact angle. Overall, the contaminated droplet demonstrates large deformation than that of the clean droplet. In other words, the surfactants make the droplet more ``active'' to the forcing movement and thus helps dewetting. This property is often used in the oil exploitation.

\section{Conclusion}
\label{sec:con}

In this work, we considered the dynamic of a droplet on the solid substrate with soluble surfactants in the droplet. The soluble surfactants consist of a bulk part dissolving inside the droplet and a interfacial part adsorbed on the fluid interface. The total free energy includes the kinetic energy, the chemical energy in the droplet due to the addition of bulk surfactants, and the surface free energies of the fluid-fluid and the fluid-solid interfaces.

We developed a thermodynamically consistent sharp-interface model for this problem, which is a coupled system of the two-phase Navier-Stokes equations for the fluid dynamic and bulk\slash surface convection-diffusion equations for the surfactant transport. The boundary\slash interface conditions and the surfactant adsorption\slash desorption condition were derived to guarantee that the total free energy decays with time. For different forms of the surface free energy, we recovered various classical adsorption isotherms and identified the adsorption\slash desorption coefficients from other phenomenological parameters.

We proposed an Eulerian weak formulation for the Navier-Stokes equation and an arbitrary Lagrangian-Eulerian weak formulation for the surfactant transport equations. The moving mesh approach was used so that the evolving fluid interface remains fitted to the triangular mesh in the bulk. We discretized the two weak formulations to obtain the finite element approximations on the moving mesh. The resulting numerical method is shown to conserve the total mass of the surfactants exactly. We carried out numerical simulations to show the convergence and accuracy of the numerical method. We also numerically examined the surfactant influence on the contact line dynamics during the droplet spreading and migration. The dependencies of the equilibrium contact angle on the dimensionless adsorption parameters were quantitatively studied.

Although the numerical study in the current work focused on two-dimensional moving contact line problems with Langmuir adsorption isotherm, the proposed numerical method can be generalized to three-dimensional problems. In the future work, we intend to numerically study the three-dimensional moving contact line problems with soluble surfactants. We will also investigate moving contact line problems with other adsorption isotherms. Some interesting problems, such as the droplet impact on solid substrates and the contact angle hysteresis in the presence of the surfactants, will also be our future concern.

\appendix

\section{Differential calculus}\label{sec:appA}
Consider a function $\zeta(\vec x, t)$ defined over the moving bulk domain $\Omega_1(t)\subset\mathbb{R}^d$ with $d=2,3$, which can be regarded as the image of a reference Lagrangian domain $\mathcal{D}\subset\mathbb{R}^d$ through a family of mappings $\{\mathcal{B}_t\}_{t\in[0,T]}$, i.e., $\mathcal{B}_t(\bmath{\xi})=\vec x(\bmath{\xi},~t)$, where $\bmath{\xi}\in\mathcal{D},\vec x\in\Omega_1(t)$. We further assume $\mathcal{B}_t\in \left[W^{1,\infty}(\mathcal{D})\right]^2,\, \mathcal{B}_t^{-1}\in\left[W^{1,\infty}(\Omega_1(t))\right]^2,\, \forall t\in[0,~T]$. Then the velocity field of the domain $\Omega_1(t)$ is given by
\begin{align*}
\vec u(\vec x,~t)= \frac{\partial\vec x(\bmath{\xi},~t)}{\partial t}=\frac{\partial\vec x}{\partial t}(\mathcal{B}_t^{-1}(\vec x),~t).
\end{align*}
The material derivative is defined as the time derivative with respect to the Lagrange frame $\mathcal{D}$ and is calculated in the following way:
\begin{align}
\label{eqn:mderivative}
\frac{\rD\zeta}{\rD t} :=\left.\frac{\partial\zeta}{\partial t}\right|_{\mathcal{D}}=\frac{\partial(\zeta\circ\mathcal{B}_t)}{\partial t}(\bmath{\xi},~t)= \frac{\partial\zeta}{\partial t} + \vec u\cdot\nabla\zeta.
\end{align}
The Reynolds transport formula on the bulk domain states that
\begin{align}
\label{eqn:dderivative}
\frac{{\rm d}}{\dt}\int_{\Omega_1(t)} \zeta(\vec x,~t)\,\dH^d=\int_{\Omega_1(t)}\left(\frac{{\rD}\zeta}{\rD t} + \zeta\,\nabla\cdot\vec u\right)\,\dH^d.
\end{align}
Given a function $\phi(\vec x, ~t)$ defined on the $(d-1)$-dimensional moving manifold $\Sigma(t)$, the Reynolds transport formula on $\Sigma(t)$ states that
\begin{align}
\label{eqn:sderivative}
\frac{{\rm d}}{\dt}\int_{\Sigma(t)} \phi(\vec x,~t)\,\dH^{d-1}=\int_{\Sigma(t)}\left(\frac{\rD\phi}{\rD t} + \phi\,\nabla_s\cdot\vec u\right)\,\dH^{d-1},
\end{align}
where $\nabla_s=(\mathbf{I}-\bn\otimes\bn)\nabla=\mathcal{P}_{_\Sigma}\nabla$ is the surface gradient operator with $\mathcal{P}_{_\Sigma}$ being the projection operator onto $\Sigma$.

Finally, we have the formula for the integration by parts on $\Sigma$ as
 \begin{align}
 \label{eqn:integrationbyparts}
 \int_{\Sigma}\nabla_s\phi\,\dH^{d-1} = \int_{\Sigma}\phi\,\kappa\,\vec n\,\dH^{d-1}  + \int_{\Lambda}\phi\,\mathbf{m}\,\dH^{d-2},
 \end{align}
where $\kappa=\nabla_s\cdot\vec n$ is the mean curvature of $\Sigma$, and $\mathbf{m}$ the co-normal vector of $\Sigma$ at $\Lambda$ (see Fig.~\ref{Fig:dynamicmodel}). Similarly, the integration by parts formula for a smooth vector field $\vec F$ reads
\begin{align}
 \label{eqn:integrationbyparts1}
 \int_{\Sigma}\nabla_s\cdot\vec F\,\dH^{d-1} = \int_{\Sigma}\vec F\cdot\vec n\,\kappa\,\dH^{d-1}  + \int_{\Lambda}\vec F\cdot\mathbf{m}\,\dH^{d-2}.
 \end{align}
 By using the identity $\nabla_s\cdot(\phi\vec F) = \nabla_s\phi\cdot\vec F + \phi\nabla_s\cdot\vec F$, we have
 \begin{align}
 \label{eqn:integrationbyparts2}
 \int_{\Sigma}\phi\nabla_s\cdot\vec F\,\dH^{d-1} = -\int_{\Sigma}\nabla_s\phi\cdot\vec F + \int_{\Sigma}\phi\,\vec F\cdot\vec n\,\kappa\,\dH^{d-1}  + \int_{\Lambda}\phi\,\vec F\cdot\mathbf{m}\,\dH^{d-2}.
 \end{align}

\section*{Acknowledgement}
We are grateful to Tiezheng Qian (The Hong Kong University of Science and Technology) and Chun Liu (Illinois Institute of Technology, Chicago) for helpful discussions.
The work of Ren was partially supported by Singapore MOE RSB grant, Singapore MOE AcRF grants (No.
R-146-000-267-114, No. R-146-000-285-114) and NSFC grant (No. 11871365). The work of Zhang was partially supported by the NSFC grant
(No. 11731006, No. 12071207) and the Guangdong Provincial Key Laboratory of Computational Science and Material Design (No. 2019B030301001).

\bibliographystyle{elsarticle-num}
\bibliography{thebib}
\end{document}